\documentclass[10pt, twocolumn,article]{IEEEtran}
\usepackage{setspace}
\usepackage{cite}
\usepackage[latin1]{inputenc}
\usepackage{epsf}\epsfverbosetrue
\usepackage{graphics,epsfig,color}
\usepackage{graphicx,epsfig}
\usepackage{epsfig}
\usepackage{multirow}
\usepackage{slashbox}
\usepackage{alltt}
\usepackage{subfigure}
\usepackage{caption}
\usepackage{color}
\usepackage{float}
\usepackage{url}
\usepackage{graphicx}
\usepackage{verbatim} 
\usepackage{footnote} 
\usepackage{latexsym} 
\usepackage{amsmath}
\usepackage{sidecap} 
\usepackage{color}
\usepackage{wrapfig}
\usepackage{algorithm}
\usepackage{eso-pic}
\usepackage{fix-cm}
\usepackage{mcite}
\usepackage{layout}
\usepackage{amsmath}
\usepackage{textcomp}
\usepackage{multirow}
\usepackage{soul}
\usepackage{amsmath}
\usepackage{verbatim}
\usepackage{bbding}

\newcommand{\ans}[1]{\textcolor{black}{#1}}
\newcommand{\hashim}[1]{\textcolor{black}{#1}}
\newcommand\tab[1][0.8cm]{\hspace*{#1}}

\begin{document}

\title{Remaining Idle Time Aware Intelligent Channel Bonding Schemes for Cognitive Radio Sensor Networks}

\author{
Syed Hashim Raza Bukhari$^{\star \diamond}$, Mubashir Husain Rehmani$^{\theta}$, and Sajid Siraj$^{\star \times}$ \vspace{2mm} \\
$^{\star}$ COMSATS Institute of Information Technology, Wah Cantt, Pakistan, $^{\diamond}$ COMSATS Institute of Information Technology, Attock, Pakistan, $^\theta$ Waterford Institute of Technology (WIT) Ireland, $^{\times}$ University of Leeds, United Kingdom  \\
          {\normalsize hashimbukhari01@gmail.com, mshrehmani@gmail.com and sajidsiraj@gmail.com}
}

\maketitle

\begin{abstract}
Channel Bonding (CB) is a technique used to provide larger bandwidth to users. It has been applied to various networks such as wireless local area networks (WLANs), Wireless Sensor Networks (WSNs), Cognitive Radio Networks (CRNs), and Cognitive Radio Sensor Networks (CRSNs). The implementation of CB in CRSNs needs special attention as Primary Radio (PR) nodes traffic must be protected from any harmful interference by Cognitive Radio (CR) sensor nodes. \ans{On the other hand, CR sensor nodes need to communicate without interruption to meet their data rate requirements and conserve energy}. If CR nodes perform frequent channel switching due to PR traffic then it will be difficult to meet their quality of service (QoS) and data rate requirements. So, CR nodes need to select those channels which are stable. \ans{By stable, we mean those channels which having less PR activity or long remaining idle time and cause less harmful interference to PR nodes.} In this paper, we propose two approaches Remaining Idle Time aware intelligent Channel Bonding (RITCB) and Remaining Idle Time aware Intelligent Channel Bonding with Interference Prevention (RITCB-IP) for cognitive radio sensor networks which select \ans{stable} channels for CB which have longest remaining idle time. We compare our approaches with four schemes \ans{such as Primary radio user activity aware channel bonding scheme (PRACB), Sample width algorithm (SWA), Cognitive radio network over white spaces (KNOWS) and AGILE}. Simulation results show that our proposed approaches RITCB and RITCB-IP decrease harmful interference (HIR) and increases the life time of cognitive radio sensor nodes.
\end{abstract}



\section{Introduction}


Wireless Sensor Networks (WSNs) play an important role in automation of our daily life processes and industries. Potential applications of WSNs includes military target tracking and surveillance \cite{Yick2008}, temperature monitoring, humidity monitoring, agriculture \cite{NingWang2006}, food health monitoring \cite{qi2014c}, vehicle movemement detection, forest fire detection, and health applications \cite{Akyildiz2002}. All these applications have made WSNs so attractive that they can be found almost everywhere. \ans{Emerging smart cities are majorly building their infrastructure on these WSNs as their vital components \cite{Bukhari2018}}. As wireless sensor nodes utilize industrial, scientific, and medical (ISM) band for communication so the problem arises when overlaid deployment of multiple WSNs is required in the same geographical area \cite{Ahmad2015}. For instance, in forests, multiple sensor networks are deployed to sense the specific parameteres. Like sensor network `A' is for smoke detection to generate a fire alarm, sensor network `B' is for humudity level detection and sensor network `C' is for animals movement detection. These different sensor networks utilize the same ISM band for their communication and are deployed in same geographical area. So, their overlaid deployment will create inteference among the co-existent sensor networks and also create spectrum scarcity and/or utilization problems \cite{Akyildiz2002}. To solve this issue of overlaid deployment and spectrum scarcity, cognitive radio capability can be added to WSNs. When wireless sensor nodes are equipped with cognitive radio capability, they are called as Cognitive Radio Sensor Networks (CRSNs). With cognitive radio support, sensor nodes can utilize the unlicensed as well as licensed band opportunistically, which improves the spectrum utilization and sensor nodes become able to meet their quality of service (QoS) and data rate requirements \cite{akhtar2016white}. This capability also helps to mitigate the spectrum scarcity problem in WSNs \cite{gulbahar2012information}. CRSNs also help to minimize the ISM band interference and jamming \cite{oto2012energy}. CRSNs have been promising in various real world scenarios such as indoor sensing applications, multimedia applications, multi class heterogeneous sensing applications and real time surveillance applications \cite{Akan2009}. The dynamic spectrum access (DSA) has powered up these CRSNs through which they can provide required performance in the diversity of applications by changing the operating parameters dynamically and adapting according to the channels condition. In CRSNs, there are two types of users which utilize the spectrum. The users which have a license to access the channel are called primary users (PR) and the unlicensed opportunistic users are called as cognitive users (CR). 

\begin{table*}[t]
\centering
\resizebox{\textwidth}{!}{%
\footnotesize
\textcolor{black}{}%
\begin{tabular}{|c|c|c|c|c|}
\hline 
\textcolor{black}{References} & \textcolor{black}{Considered Network} & \textcolor{black}{Publication Year} & \textcolor{black}{Description} \tabularnewline
\hline 
\textcolor{black}{\cite{sharma2010opportunistic}} & \multirow{6}{*}{} & \textcolor{black}{2010} & \textcolor{black}{Residual idle time distribution scheme to estimate the transmission duration in remaining idle time} \\ \cline{1-1} \cline{3-4}
\textcolor{black}{\cite{song2010common}} &  & \textcolor{black}{2010} & \textcolor{black}{A proactive scheme for unlicensed users to perform channel switching before licensed user appear on the network} \\ \cline{1-1} \cline{3-4}
\textcolor{black}{\cite{wang2009modeling}} &  & \textcolor{black}{2009} & \textcolor{black}{The need to perform intelligent channel decision for improved utilization} \\ \cline{1-1} \cline{3-4}
\textcolor{black}{\cite{li2009traffic}} & \textcolor{black}{CRNs} & \textcolor{black}{2009} & \textcolor{black}{Improved spectrum sharing based on traffic pattern prediction in cognitive radio networks} \\ \cline{1-1} \cline{3-4}
\textcolor{black}{\cite{li2008traffic}} &  & \textcolor{black}{2008} & \textcolor{black}{Forecasting the traffic pattern to reduce the frequency hopping rate and interference effects in CRNs} \\ \cline{1-1} \cline{3-4}
\textcolor{black}{\cite{L.Yang2008}} &  & \textcolor{black}{2008} & \textcolor{black}{The intelligent utilization of channel histories to predict the forthcoming PR activities} \tabularnewline
\hline 
\end{tabular}%
}

\protect\caption{Summary of Remaining Idle Time based schemes for CRNs in Literature}
 \label{tab:RIT-schemes}

\end{table*}

The PR nodes are the priority nodes and whenever CRSN node\footnote{We use the term CR nodes and CRSN nodes for the same types of nodes throughout the manuscript and they are used interchangeably.} have some data to transmit, it has to go through a three step process known as \textit{cognitive cycle} \cite{Liang2008}. It first performs channel sensing to know about the present channel condition. Based on the channel sensing results, the CRSN node take the decision to tune to or select the suitable channel for communication. The CRSN nodes can now access both licensed and unlicensed bands opportunistically provided that they will not create harmful interference to the PR nodes. If any PR node appear on the same channel during CRSN communication, the CRSN node has to immediately \hashim{leave} the channel for PR communication. The addition of CR capability to wireless sensor nodes for effective spectrum utilization has also invoked certain challenges as well. These challenges must be addressed for prevailing future of CRSNs. 

As multimedia content has become the integral part of all applications and CRSNs tend to provide seamless communication for those applications uning unlicensed opportunistic access \cite{Akan2009}, therefore, large bandwidth is required to meet data rate requirements of CRSNs. CB tends to empower CRSNs by providing large bandwidth to opportunistic users. CB proposes to take benefit of multiple contiguous channels which are free from PR activity. CB combines the multiple contiguous channels to be used by high bandwidth applications. \ans{The devices are now equipped with multiple interfaces which sense the spectrum holes in parallel for multiple networks and utilize them whenever found \cite{Lin2013}.} Although CB is a very promising approach in \ans{WLANs, WSNs, CRNs and for CRSNs}, CB can provide the required performance only when it accurately \hashim{makes} decision on the basis of PR activity. Primary Radio (PR) activity is an approximation of spectrum utilization by PR nodes. 

The channel selection is based on channel sensing process. \ans{The CRSN nodes sense the channels and if they are found free, they can utilize those channel for CB. During communication, the CRSN nodes do periodic channel sensing and if any PR node appears on any channel of bond, the CR node has to stop its communication, break the bond and leave the channel}. It means that the PR node will have to face interference before being detected by the CRSN node. Similarly, the CRSN transmissions face a great disruption \ans{due to frequent making and breaking of bonds} and it becomes highly difficult for CRSN nodes to meet the certain network requirements. \ans{Here, the importance of RIT gets highlighted. If the CRSN nodes select those channels which have highest probability of longest RIT, they can meet their QoS requirements and reduce harmful interference to PR nodes.} Hence, the channels with longest remaining idle time are suitable for CRSN nodes. In this way, the harmful interference can be reduced and CRSN nodes may also become able to meet certain QoS requirements \cite{sharma2010opportunistic}. 

In this paper, we propose two schemes RITCB and RITCB-IP for CRSNs which select those channels for CRSN nodes which have longest remaining idle time. 


\subsection{Contribution in this Article:}
Our contributions in this article are as follows:
\begin{itemize}
\item {We propose a channel bonding algorithm which selects channels based on longest remaining idle time (RITCB) for the purpose of channel bonding.}
\item {We then propose a modified scheme called as RITCB-IP to prevent harmful interference with PR nodes.}
\item {We analyze our proposed channel bonding algorithm under different PR activities and compare our two proposed schemes RITCB and RITCB-IP with PRACB, SWA, KNOWS and AGILE.}
\end{itemize}

\subsection{Article Structure:}

In the next section, i.e., \hashim{section II}, we discuss the related work. In \hashim{section III}, we discuss our proposed remaining idle time aware channel bonding algorithm for CRSNs. The performance analysis of our proposed scheme has been discussed in \hashim{section IV} and paper concludes in \hashim{section V}.
\section{Related Work}

First, we briefly discuss the use of CB in WLANs, WSNs, and CRNs. Then we discuss the challenges of CB while implementing in CRSNs.



\begin{table*}[htbp]
\begin{center}
\begin{tabular}{|l|l|l|l|l|l|l|l|l|l|l|l|l|l|l|l|}
\hline 
 & \textbf{{\tiny{}Ch 0}} & \textbf{{\tiny{}Ch 1}} & \textbf{{\tiny{}Ch 2}} & \textbf{{\tiny{}Ch 3}} & \textbf{{\tiny{}Ch 4}} & \textbf{{\tiny{}Ch 5}} & \textbf{{\tiny{}Ch 6}} & \textbf{{\tiny{}Ch 7}} & \textbf{{\tiny{}Ch 8}} & \textbf{{\tiny{}Ch 9}} & \textbf{{\tiny{}Ch 10}} & \textbf{{\tiny{}Ch 11}} & \textbf{{\tiny{}Ch 12}} & \textbf{{\tiny{}Ch 13}} & \textbf{{\tiny{}Ch 14}}\tabularnewline
\hline
\hline 
\textbf{{\tiny{}T$_{ON}$}} & {\tiny{}0.83} & {\tiny{}0.77} & {\tiny{}0.42} & {\tiny{}0.31} & {\tiny{}0.53} & {\tiny{}0.27} & {\tiny{}0.36} & {\tiny{}0.2} & {\tiny{}0.26} & {\tiny{}0.24} & {\tiny{}0.13} & {\tiny{}0.15} & {\tiny{}0.18} & {\tiny{}0.48} & {\tiny{}0.3}\tabularnewline
\hline 
\textbf{{\tiny{}T$_{OFF}$}} & {\tiny{}2.5} & {\tiny{}1.11} & {\tiny{}10.0} & {\tiny{}1.67} & {\tiny{}3.33} & {\tiny{}10.0} & {\tiny{}4.0} & {\tiny{}9.09} & {\tiny{}3.45} & {\tiny{}2.08} & {\tiny{}5.26} & {\tiny{}3.7} & {\tiny{}1.0} & {\tiny{}1.61} & {\tiny{}2.63}\tabularnewline
\hline 
\textbf{{\tiny{}$\lambda_{X}$}} & {\tiny{}1.20} & {\tiny{}1.29} & {\tiny{}2.38} & {\tiny{}3.22} & {\tiny{}1.88} & {\tiny{}3.70} & {\tiny{}2.77} & {\tiny{}5} & {\tiny{}3.84} & {\tiny{}4.16} & {\tiny{}7.69} & {\tiny{}6.66} & {\tiny{}5.55} & {\tiny{}2.08} & {\tiny{}3.33}\tabularnewline
\hline 
\textbf{{\tiny{}$\lambda_{Y}$}} & {\tiny{}0.4} & {\tiny{}0.90} & {\tiny{}0.1} & {\tiny{}0.59} & {\tiny{}0.30} & {\tiny{}0.1} & {\tiny{}0.25} & {\tiny{}0.11} & {\tiny{}0.28} & {\tiny{}0.48} & {\tiny{}0.19} & {\tiny{}0.27} & {\tiny{}1} & {\tiny{}0.62} & {\tiny{}0.38}\tabularnewline
\hline 
\textbf{{\tiny{}$\mu^{i}$}} & {\tiny{}0.24} & {\tiny{}0.40} & {\tiny{}0.04} & {\tiny{}0.15} & {\tiny{}0.13} & {\tiny{}0.02} & {\tiny{}0.08} & {\tiny{}0.02} & {\tiny{}0.07} & {\tiny{}0.10} & {\tiny{}0.02} & {\tiny{}0.03} & {\tiny{}0.15} & {\tiny{}0.22} & {\tiny{}0.10}\tabularnewline
\hline
\end{tabular}%
\end{center}
\protect\caption{Wireless Channel Parameters used in simulation (Low PR Activity)}
\label{Wireless-channel-Parameters-low}

\end{table*}

\begin{table*}[htbp]
\begin{center}
\begin{tabular}{|l|l|l|l|l|l|l|l|l|l|l|l|l|l|l|l|}
\hline 
 & \textbf{{\tiny{}Ch 0}} & \textbf{{\tiny{}Ch 1}} & \textbf{{\tiny{}Ch 2}} & \textbf{{\tiny{}Ch 3}} & \textbf{{\tiny{}Ch 4}} & \textbf{{\tiny{}Ch 5}} & \textbf{{\tiny{}Ch 6}} & \textbf{{\tiny{}Ch 7}} & \textbf{{\tiny{}Ch 8}} & \textbf{{\tiny{}Ch 9}} & \textbf{{\tiny{}Ch 10}} & \textbf{{\tiny{}Ch 11}} & \textbf{{\tiny{}Ch 12}} & \textbf{{\tiny{}Ch 13}} & \textbf{{\tiny{}Ch 14}}\tabularnewline
\hline 
\hline
\textbf{{\tiny{}T$_{ON}$}} & {\tiny{}3.33} & {\tiny{}1.11} & {\tiny{}10.0} & {\tiny{}5.0} & {\tiny{}2.5} & {\tiny{}1.67} & {\tiny{}2.86} & {\tiny{}5.56} & {\tiny{}5.88} & {\tiny{}4.35} & {\tiny{}1.85} & {\tiny{}1.3} & {\tiny{}1.0} & {\tiny{}1.23} & {\tiny{}2.38}\tabularnewline
\hline 
\textbf{{\tiny{}T$_{OFF}$}} & {\tiny{}0.83} & {\tiny{}0.77} & {\tiny{}0.42} & {\tiny{}0.31} & {\tiny{}0.53} & {\tiny{}0.27} & {\tiny{}0.36} & {\tiny{}0.2} & {\tiny{}0.26} & {\tiny{}0.24} & {\tiny{}0.13} & {\tiny{}0.15} & {\tiny{}0.18} & {\tiny{}0.48} & {\tiny{}0.3}\tabularnewline
\hline 
\textbf{{\tiny{}$\lambda_{X}$}} & {\tiny{}0.30} & {\tiny{}0.90} & {\tiny{}0.1} & {\tiny{}0.2} & {\tiny{}0.4} & {\tiny{}0.59} & {\tiny{}0.34} & {\tiny{}0.17} & {\tiny{}0.17} & {\tiny{}0.22} & {\tiny{}0.54} & {\tiny{}0.76} & {\tiny{}1} & {\tiny{}0.81} & {\tiny{}0.42}\tabularnewline
\hline 
\textbf{{\tiny{}$\lambda_{Y}$}} & {\tiny{}1.20} & {\tiny{}1.29} & {\tiny{}2.38} & {\tiny{}3.22} & {\tiny{}1.88} & {\tiny{}3.70} & {\tiny{}2.77} & {\tiny{}5} & {\tiny{}3.84} & {\tiny{}4.16} & {\tiny{}7.69} & {\tiny{}6.66} & {\tiny{}5.55} & {\tiny{}2.08} & {\tiny{}3.33}\tabularnewline
\hline 
\textbf{{\tiny{}$\mu^{i}$}} & {\tiny{}0.80} & {\tiny{}0.59} & {\tiny{}0.95} & {\tiny{}0.94} & {\tiny{}0.82} & {\tiny{}0.86} & {\tiny{}0.88} & {\tiny{}0.96} & {\tiny{}0.95} & {\tiny{}0.94} & {\tiny{}0.93} & {\tiny{}0.89} & {\tiny{}0.84} & {\tiny{}0.71} & {\tiny{}0.88}\tabularnewline
\hline
\end{tabular}%
\protect\caption{Wireless Channel Parameters used in simulation (High PR Activity)}
\label{Wireless-channel-Parameters-high}

\end{center}
\end{table*}

\subsection{RIT based CB in WLANs and WSNs}

CB combines the multiple contiguous channels to provide the users a larger bandwidth \cite{bukhari2016}. WLANs are now being deployed for providing continuous seamless support round the clock to the users. Almost all the wireless devices are powered by WLAN support from their vendors. Due to this facility, the users can get internet support almost every where either in the home, office, hospital or market etc. As, WLANs are mostly used for providing internet access to the users, hence the performance can be enhanced by performing CB. The new standard for WLAN IEEE 802.11n supports CB to provide large bandwidth to the users. Hence, all the users in WLAN environment are sharing the sources equally so, the size of bond will vary according to the number of users present at a time. If there are more users present and require network resources, the bond size will decrease to facilitate more users and if less users are present, the bond size can be increased. The several ways to improve bandwidth in wireless networks have been discussed in \cite{Ramaboli2012} in which authors have highlighted the effectiveness of using CB approach.

WSNs are deployed to sense some specific event and transmit its parameters in a multi-hop manner to the central station or actuator \cite{Akyildiz2002}. WSNs are normally deployed densely as they have specific communication range and lifetime. Using short communication range, the life time of wireless sensor nodes can be increased. The current standard for WSNs is IEEE 802.15.4 does not \hashim{support} CB by default. As, WSNs normally utilize industrial, scientific and medical (ISM) band for communication, so, effective spectrum utilization is necessary to avoid any interference.  

\subsection{RIT based CB in CRNs}

CRNs are opportunistic nodes which utilize ISM as well as licensed band for their communication. As in a CR based network, CR nodes co-exist with the licensed nodes so it is required that CR nodes must not create interference for the licensed users. It is possible that a CR node is transmitting its data and meanwhile a PR nodes tries to access the same frequency band for transmission. In this situation, the CR node must leave the band by stopping its communication to prevent from PR-CR interference. The information of PR activity is necessary for interference prevention. In this regard, PR activity model is required to estimate the presence or absence of PR node. 

The CR technique helps to improve the utilization of scarce spectrum. The CR nodes in a network utilize the spectrum opportunities (white spaces) while avoiding harmful interference. The intellient decision provides CR node a better chance to exploit the opportunities. The authors in \cite{wang2009modeling} have mentioned that intelligent channel utilization is dependent upon total number of CR nodes in a network. If the CR traffic is low in network then probability based channel selection can provide required performance but for high CR traffic, sensing based channel selection provides better results to select the required channels. \cite{li2009traffic} highlights the importance of channel prediction techniques for CR based networks. Multiple methods have been introduced to estimate the suitability of channels for CR nodes. Using these methods, CR nodes can switch to those channels which provide longer idle slots through network coordination \cite{song2010common}. Furthermore \cite{li2008traffic} discusses that prediction based schemes for PR users activity can enhance the CR node performance as it reduces the number of channels switching. The CR nodes can intelligently utilize the past channel histories to predict the forthcoming PR activities and can effectively reduce the harmful interference \cite{L.Yang2008}. When CR capabilities are addded to WSN node, the power consumption also becomes an important parameter. An effective way of increasing the life of CR based wireless sensor node is to perform channel sensing only when a CR node has some data to send \cite{sharma2010opportunistic, sharma2010residual}, thus the battery resources can be saved remarkably. 

CB in CR based networks provides larger bandwidth to CR nodes while keeping the interference below the threshold level. The emerging CR based standard IEEE 802.22 supports CB to provide larger bandwidth to CR nodes \cite{Cordeiro2006a}. 

The CB in CRSNs is a new idea and we have proposed it in \cite{bukhari2016pracb}. The work shows that CB in CRSNs using efficient channel selection can reduce harmful interference. \textit{Now, in this paper, we propose two channel bonding algorithms based on RIT. Our work is different from all the aforementioned literature of CR networks in a sense that we consider CRSN while the works presented in the literature are specifically designed for CRNs (see Table. \ref{tab:RIT-schemes}) and these works cannot be directly applied to CRSNs.} 

\begin{table*}[htbp]
\begin{center}

\begin{tabular}{|l|l|l|l|l|l|l|l|l|l|l|l|l|l|l|l|}
\hline 
 & \textbf{{\tiny{}Ch 0}} & \textbf{{\tiny{}Ch 1}} & \textbf{{\tiny{}Ch 2}} & \textbf{{\tiny{}Ch 3}} & \textbf{{\tiny{}Ch 4}} & \textbf{{\tiny{}Ch 5}} & \textbf{{\tiny{}Ch 6}} & \textbf{{\tiny{}Ch 7}} & \textbf{{\tiny{}Ch 8}} & \textbf{{\tiny{}Ch 9}} & \textbf{{\tiny{}Ch 10}} & \textbf{{\tiny{}Ch 11}} & \textbf{{\tiny{}Ch 12}} & \textbf{{\tiny{}Ch 13}} & \textbf{{\tiny{}Ch 14}}\tabularnewline
\hline
\hline 
\textbf{{\tiny{}T$_{ON}$}} & {\tiny{}3.33} & {\tiny{}1.11} & {\tiny{}10.0} & {\tiny{}5.0} & {\tiny{}2.5} & {\tiny{}1.67} & {\tiny{}2.86} & {\tiny{}5.56} & {\tiny{}5.88} & {\tiny{}4.35} & {\tiny{}1.85} & {\tiny{}1.3} & {\tiny{}1.0} & {\tiny{}1.23} & {\tiny{}2.38}\tabularnewline
\hline 
\textbf{{\tiny{}T$_{OFF}$}} & {\tiny{}2.5} & {\tiny{}1.11} & {\tiny{}10.0} & {\tiny{}1.67} & {\tiny{}3.33} & {\tiny{}10.0} & {\tiny{}4.0} & {\tiny{}9.09} & {\tiny{}3.45} & {\tiny{}2.08} & {\tiny{}5.26} & {\tiny{}3.7} & {\tiny{}1.0} & {\tiny{}1.61} & {\tiny{}2.63}\tabularnewline
\hline 
\textbf{{\tiny{}$\lambda_{X}$}} & {\tiny{}0.30} & {\tiny{}0.90} & {\tiny{}0.1} & {\tiny{}0.2} & {\tiny{}0.4} & {\tiny{}0.59} & {\tiny{}0.34} & {\tiny{}0.17} & {\tiny{}0.17} & {\tiny{}0.22} & {\tiny{}0.54} & {\tiny{}0.76} & {\tiny{}1} & {\tiny{}0.81} & {\tiny{}0.42}\tabularnewline
\hline 
\textbf{{\tiny{}$\lambda_{Y}$}} & {\tiny{}0.4} & {\tiny{}0.90} & {\tiny{}0.1} & {\tiny{}0.59} & {\tiny{}0.30} & {\tiny{}0.1} & {\tiny{}0.25} & {\tiny{}0.11} & {\tiny{}0.28} & {\tiny{}0.48} & {\tiny{}0.19} & {\tiny{}0.27} & {\tiny{}1} & {\tiny{}0.62} & {\tiny{}0.38}\tabularnewline
\hline 
\textbf{{\tiny{}$\mu^{i}$}} & {\tiny{}0.57} & {\tiny{}0.5} & {\tiny{}0.5} & {\tiny{}0.74} & {\tiny{}0.42} & {\tiny{}0.14} & {\tiny{}0.41} & {\tiny{}0.37} & {\tiny{}0.63} & {\tiny{}0.67} & {\tiny{}0.26} & {\tiny{}0.26} & {\tiny{}0.5} & {\tiny{}0.43} & {\tiny{}0.47}\tabularnewline
\hline 
\end{tabular}%
\protect\caption{Wireless Channel Parameters used in simulation (Long PR Activity)}

\label{Wireless-channel-Parameters-long}

\end{center}
\end{table*}

\begin{table*}[htbp]
\begin{center}

\begin{tabular}{|l|l|l|l|l|l|l|l|l|l|l|l|l|l|l|l|}
\hline 
 & \textbf{{\tiny{}Ch 0}} & \textbf{{\tiny{}Ch 1}} & \textbf{{\tiny{}Ch 2}} & \textbf{{\tiny{}Ch 3}} & \textbf{{\tiny{}Ch 4}} & \textbf{{\tiny{}Ch 5}} & \textbf{{\tiny{}Ch 6}} & \textbf{{\tiny{}Ch 7}} & \textbf{{\tiny{}Ch 8}} & \textbf{{\tiny{}Ch 9}} & \textbf{{\tiny{}Ch 10}} & \textbf{{\tiny{}Ch 11}} & \textbf{{\tiny{}Ch 12}} & \textbf{{\tiny{}Ch 13}} & \textbf{{\tiny{}Ch 14}}\tabularnewline
\hline 
\hline 
\textbf{{\tiny{}T$_{ON}$}} & {\tiny{}0.83} & {\tiny{}0.77} & {\tiny{}0.42} & {\tiny{}0.31} & {\tiny{}0.53} & {\tiny{}0.27} & {\tiny{}0.36} & {\tiny{}0.2} & {\tiny{}0.26} & {\tiny{}0.24} & {\tiny{}0.13} & {\tiny{}0.15} & {\tiny{}0.18} & {\tiny{}0.48} & {\tiny{}0.3}\tabularnewline
\hline 
\textbf{{\tiny{}T$_{OFF}$}} & {\tiny{}0.27} & {\tiny{}0.36} & {\tiny{}0.2} & {\tiny{}0.26} & {\tiny{}0.24} & {\tiny{}0.83} & {\tiny{}0.77} & {\tiny{}0.42} & {\tiny{}0.31} & {\tiny{}0.53} & {\tiny{}0.4} & {\tiny{}0.29} & {\tiny{}0.15} & {\tiny{}0.53} & {\tiny{}0.2}\tabularnewline
\hline 
\textbf{{\tiny{}$\lambda_{X}$}} & {\tiny{}1.20} & {\tiny{}1.29} & {\tiny{}2.38} & {\tiny{}3.22} & {\tiny{}1.88} & {\tiny{}3.70} & {\tiny{}2.77} & {\tiny{}5} & {\tiny{}3.84} & {\tiny{}4.16} & {\tiny{}7.69} & {\tiny{}6.66} & {\tiny{}5.55} & {\tiny{}2.08} & {\tiny{}3.33}\tabularnewline
\hline 
\textbf{{\tiny{}$\lambda_{Y}$}} & {\tiny{}3.70} & {\tiny{}2.77} & {\tiny{}5} & {\tiny{}3.84} & {\tiny{}4.16} & {\tiny{}1.20} & {\tiny{}1.29} & {\tiny{}2.38} & {\tiny{}3.22} & {\tiny{}1.88} & {\tiny{}2.5} & {\tiny{}3.44} & {\tiny{}6.66} & {\tiny{}1.88} & {\tiny{}5}\tabularnewline
\hline 
\textbf{{\tiny{}$\mu^{i}$}} & {\tiny{}0.75} & {\tiny{}0.68} & {\tiny{}0.67} & {\tiny{}0.54} & {\tiny{}0.68} & {\tiny{}0.24} & {\tiny{}0.31} & {\tiny{}0.32} & {\tiny{}0.45} & {\tiny{}0.31} & {\tiny{}0.24} & {\tiny{}0.34} & {\tiny{}0.54} & {\tiny{}0.47} & {\tiny{}0.6}\tabularnewline
\hline 
\end{tabular}%
\protect\caption{Wireless Channel Parameters used in simulation (Intermittent PR Activity)}
\label{Wireless-channel-Parameters-inter}

\end{center}
\end{table*}

\section{Remaining Idle Time Aware Channel Bonding Scheme}

\subsection{System Model:}

PR activity gives the information about presence or absence of PR users over the channel. We modeled the PR activity as continuous-time, alternating ON/OFF Markov Renewal Process (MRP) \cite{G.Yuan2010,Min2008,YasirSaleem2014}. This PR activity model has been widely used in the literature \cite{Min2008,G.Yuan2010,Kim2008a,Mehanna2009,Zahmati2010,L.Yang2008}. The ON/OFF PR activity model approximates the spectrum utilization pattern of voice networks \cite{Adas1997} and also very famous for public safety bands \cite{L.Yang2008,Vujicic2005}. \ans{The presence of PR node in a wireless channel can be modelled as the ON state which represents that channel is currently busy and occupied by PR node.} The OFF state represents that channel is idle and unoccupied by any PR node. The time duration for which the channel $i$ is in ON and OFF states are denoted as $T_{ON}^{i}$ and $T_{OFF}^{i}$ respectively. The duration which a channel takes to complete one consecutive ON and OFF period is called renewal period. \ans{Let,} this renewal period for a channel $i$ at time $t$ is denoted by $Z_{i}(t)=T_{ON}^{i}+T_{OFF}^{i}$ \cite{Min2008,Kim2008,Sriram1986}. Both ON and OFF periods are assumed to be independent and identically distributed (i.i.d). Since each PR user arrival is independent so according to \cite{Sriram1986}, each PR user arrival follows the Poisson arrival process and $\lambda_{X}$ and $\lambda_{Y}$ are the rate parameters for exponential distribution, the length of ON periods which are exponentially distributed with \ans{probability density function} p.d.f. can be given as:
\begin{equation}
f_{X}(t)=\lambda_{X}\times e^{-\lambda_{X}t}
\end{equation}

Similarly, the length of OFF periods exponentially distributed with p.d.f can be stated as:
\begin{equation}
f_{Y}(t)=\lambda_{Y}\times e^{-\lambda_{Y}t}
\end{equation} 

The total utilization of channel $i$ by a PR user is called as utilization factor of $i^{th}$ channel and can be written as:  

\begin{equation}
u^{i}=\frac{E[T_{ON}^{i}]}{E[T_{ON}^{i}]+E[T_{OFF}^{i}]}=\frac{\lambda_{Y}}{\lambda_{X}+\lambda_{Y}}
\end{equation}

$E[T_{ON}^{i}]$ and $E[T_{OFF} ^{i}]$ is the mean of exponential distribution \cite{Kim2008}. $\mu_X$ and $\mu_Y$ are the mean of rate parameters where $\mu_X = \frac{1}{\lambda_{X}}$ and $\mu_Y = \frac{1}{\lambda_{Y}}$. In this way, any kind of PR activity can be added by describing  the pattern as discussed in \cite{Rehmani2013}. We have used four types of PR activities i.e. Low, High, Long and Intermittent. The wireless parameters for these four types of PR activities has been shown as Tables. \ref{Wireless-channel-Parameters-low}, \ref{Wireless-channel-Parameters-high}, \ref{Wireless-channel-Parameters-long} and \ref{Wireless-channel-Parameters-inter}.

The CRSN nodes perform spectrum sensing through detecting energy threshold on the channel. Energy threshold is the most common way of sensing due to its low complexity, computational overhead and power requirements \cite{Yucek2009}. Let $F$ is the total number of \hashim{channels which can be sensed} and $F=f_{i}$ where $i=1,2,...,m$. \ans{`$m$' is the maximum number of channels in a network}. Now $f_{i}=k(i) + n(i)$ where $k(i)$ is the signal to be detected and $n(i)$ is the additive white gaussian noise (AWGN) over the channel. Then the set of occupied channels $F_{o}$ can be formulated as:

\begin{equation}
F_{o} = f_{i} : \phi(f_{i})\geq\theta
\end{equation}

Where `$\theta$' is the threshold to take decision of a sensing function and $\phi(f_{i})$ is the sensing function. Similarly the set of vacant channels $F_{v}$ can also be formulated as:

\begin{equation}
F_{v} = f_{i} : \phi(f_{i})\leq\theta
\end{equation}

Eq. 4 and 5 provide the state of channel being occupied or vacant at any instant. We can also use probability theory to estimate the state of channel using the rate parameters. \ans{when we start our simulation, we assume that all channels are idle and calculate the probabilities for the channels to be in ON or OFF state depending upon the rate parameters.}

Let, $P_{ON}^{i}(t)$ be the probability of $i^{th}$ channel in ON state at time ``t'' and $P_{OFF}^{i}(t)$ be the probability of $i^{th}$ channel in OFF state at time ``t'' then these probabilities can be given as:

\begin{equation}
P_{ON}^{i}(t) = \frac{\lambda_{Y}^{i}}{\lambda_{X}^{i}+\lambda_{Y}^{i}} - \frac{\lambda_{Y}^{i}}{\lambda_{X}^{i}+\lambda_{Y}^{i}}.e^{-(\lambda_{X}^{i}+\lambda_{Y}^{i})t}
\end{equation}

\begin{equation}
P_{OFF}^{i}(t) = \frac{\lambda_{X}^{i}}{\lambda_{X}^{i}+\lambda_{Y}^{i}} + \frac{\lambda_{Y}^{i}}{\lambda_{X}^{i}+\lambda_{Y}^{i}}.e^{-(\lambda_{X}^{i}+\lambda_{Y}^{i})t}
\end{equation}

such that,

\begin{equation}
P_{ON}^{i}(t) + P_{OFF}^{i}(t) = 1
\end{equation}

Eq. 6 and Eq. 7 can also be written as \cite{Kim2008}:

\begin{equation}
P_{ON}^{i}(t) = u^i - u^i.e^{-(\lambda_{X}^{i}+\lambda_{Y}^{i})t}
\end{equation}

and 

\begin{equation}
P_{OFF}^{i}(t) = (1 - u^i) + u^i.e^{-(\lambda_{X}^{i}+\lambda_{Y}^{i})t}
\end{equation}

Now, following the concept of proactive planning as given in \cite{L.Yang2008}, we can use the probability of $i^{th}$ channel in the OFF state and the rate parameters to estimate the remaining idle time $RIT$ as:

\begin{equation}
RIT^{i} = \frac{P_{OFF}^{i}(t)}{\lambda_{X}^{i}}
\end{equation}

which comes to 

\begin{equation}
RIT^{i} = \frac{{\lambda_{X}^{i}+\lambda_{Y}^{i}}.e^{-(\lambda_{X}^{i}+\lambda_{Y}^{i})t}}{\lambda_{X}^{i}(\lambda_{X}^{i}+\lambda_{Y}^{i})}
\end{equation}

and 

\begin{equation}
RIT^{i} = \mu_{X}^{i}.(\frac{{\lambda_{X}^{i}+\lambda_{Y}^{i}}.e^{-(\lambda_{X}^{i}+\lambda_{Y}^{i})t}}{\lambda_{X}^{i}+\lambda_{Y}^{i}})
\end{equation}

\ans{where $P_{OFF}^{i}(t)$ is probability of node $i$ in off state at time $t$, $\lambda_{X}^{i}$ is the rate parameter and $\mu_{X}^{i}$ is the mean of rate parameter.} By using Eq. 13, we can find the RIT for channel $i$.

\begin{algorithm}[t]
  \caption{\footnotesize{Remaining Idle Time Aware Channel Bonding Algorithm for CRSNs (RITCB)}}\label{Alg}
\scriptsize
1: $CH$: Total number of channels\\
2: $CBsize$: Size for channel bonding\\
3: $SCH$: Selected channels\\
4: $CCH$: Contiguous channels\\ 
5: $RIT_i$: Remaining idle time of $i^{th}$ channel\\
6: $SRIT$: Smallest remaining idle time\\
7: $RITCH3$: The array of three contiguous channels with RIT\\
8: $RITCH2$: The array of two contiguous channels with RIT\\
9: \textbf{Start}\\
10: Perform spectrum sensing to find total number of channels $CH$\\
11: Input $CH$, $CBsize$\\
12: Select all the possible channel combinations of $CBsize$ from $CH$\\
13:	selected\_channels $\leftarrow$ add the value of possible combinations\\
14: $SCH$ $\leftarrow$ arrange the selected channels in descending order\\ 
15:	goto algorithm \ref{Alg-2} $\leftarrow$ to find set of contiguous channels\\
16:	\tab\textbf{for} ${i=0, i < CCH, i++}$ \\
17:		\tab\tab $CONT[CCH]$ $\leftarrow$ Arrange the contiguous channel pairs in descending order\\
18: \tab \tab Perform detailed channel sensing on contiguous channels to find PR activity on $CONT[CCH]$\\
19:		\tab\tab $CONT[CCH3]$ $\leftarrow$ Insert the sets of three contiguous pairs from $CONT[CCH]$\\
20:		\tab\tab $CONT[CCH2]$ $\leftarrow$ Insert the sets of two contiguous pairs from $CONT[CCH]$\\
21:	\tab\textbf{end for}	\\	
22:	 \tab\textbf{for} ${j=0, j < CONT[CCH3], j++}$ \\
23:		\tab\tab For All $j$ Compute $RIT$\\ 
24:		\tab\tab $SRIT[CCH3]$ $\leftarrow$ $min$ [$RIT(CONT[CCH3])$]\\
25:	 \tab\textbf{end for} \\
26: \tab $RITCH3$ $\leftarrow$ Select the channels from $SRIT[CCH3]$ with longest $RIT$\\
27:	 \tab\textbf{for} ${k=0, k < CONT[CCH2], k++}$ \\
28:		\tab\tab For All $k$ Compute $RIT$\\ 
29:		\tab\tab $SRIT[CCH2]$ $\leftarrow$ $min$ [$RIT(CONT[CCH2])$]\\
30:	 \tab\textbf{end for} \\
31: \tab $RITCH2$ $\leftarrow$ Select the channels from $SRIT[CCH2]$ with longest $RIT$\\
32:	\tab\textbf{if} $RITCH3$ $>$ $RITCH2$ \textbf{then}\\		
33:		\tab\tab Make a bond using three contiguous $RITCH3$\\
34:		\tab\tab Transmit data\\
35:		\tab\tab Break the bond and release the channels\\
36:	\tab\textbf{else if} $RITCH3$ $==$ $RITCH2$ \textbf{then}\\
37:		\tab\tab Make a bond using three contiguous $RITCH3$\\
38:		\tab\tab Transmit data\\
39:		\tab\tab Break the bond and release the channels\\
40:	\tab\textbf{else} $RITCH3$ $<$ $RITCH2$ \textbf{then}\\
41:		\tab\tab Make a bond using two contiguous $RITCH2$\\
42:		\tab\tab Transmit data\\
43:		\tab\tab Break the bond and release the channels\\
44: \textbf{Stop}\\

\end{algorithm}

\begin{algorithm}[t]
\caption{\footnotesize{Algorithm to find Contiguous Channels for RITCB}}\label{Alg-2}
\scriptsize
1: \textbf{Start}\\
2: Input $SCH$\\
3: For each $SCH$ pair, get the channel ID's\\
4: \tab \textbf{if} all three channels in $SCH$ are consecutive \textbf{then}\\
5:		\tab \tab $CCH$ $\leftarrow$ all selected channels are contiguous\\
6:		\tab \tab shortlist the ID's of three contiguous channels\\
7:		\tab \textbf{else if} any two channels in $SCH$ are consecutive \textbf{then}\\
8:		\tab \tab $CCH$ $\leftarrow$ two selected channels are contiguous\\
9:		\tab \tab shortlist the ID's of two contiguous channels\\
10: \textbf{Stop}\\
11: goto Algortithm \ref{Alg}\\
\end{algorithm}

\begin{figure*}
\begin{center}
\begin{centering}

\subfigure[]
{
\label{fig:wrong-selection}
	\epsfxsize= 8cm
	  \leavevmode\epsfbox{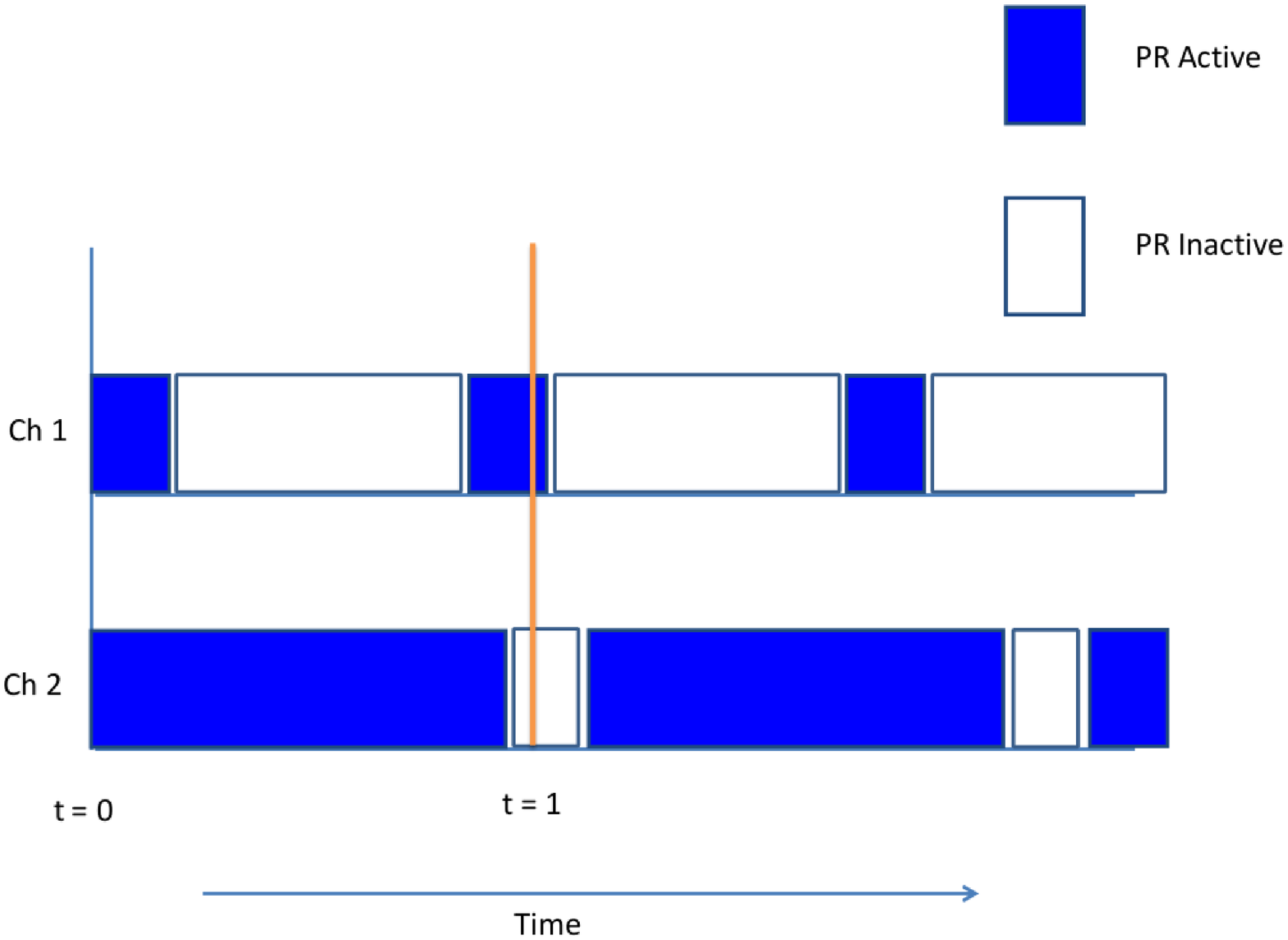}	  
}\hspace{-0.5cm}
\vspace{-0.3cm}
\subfigure[]
{
\label{fig:wrong-selection-3}
	\epsfxsize= 8cm
	  \leavevmode\epsfbox{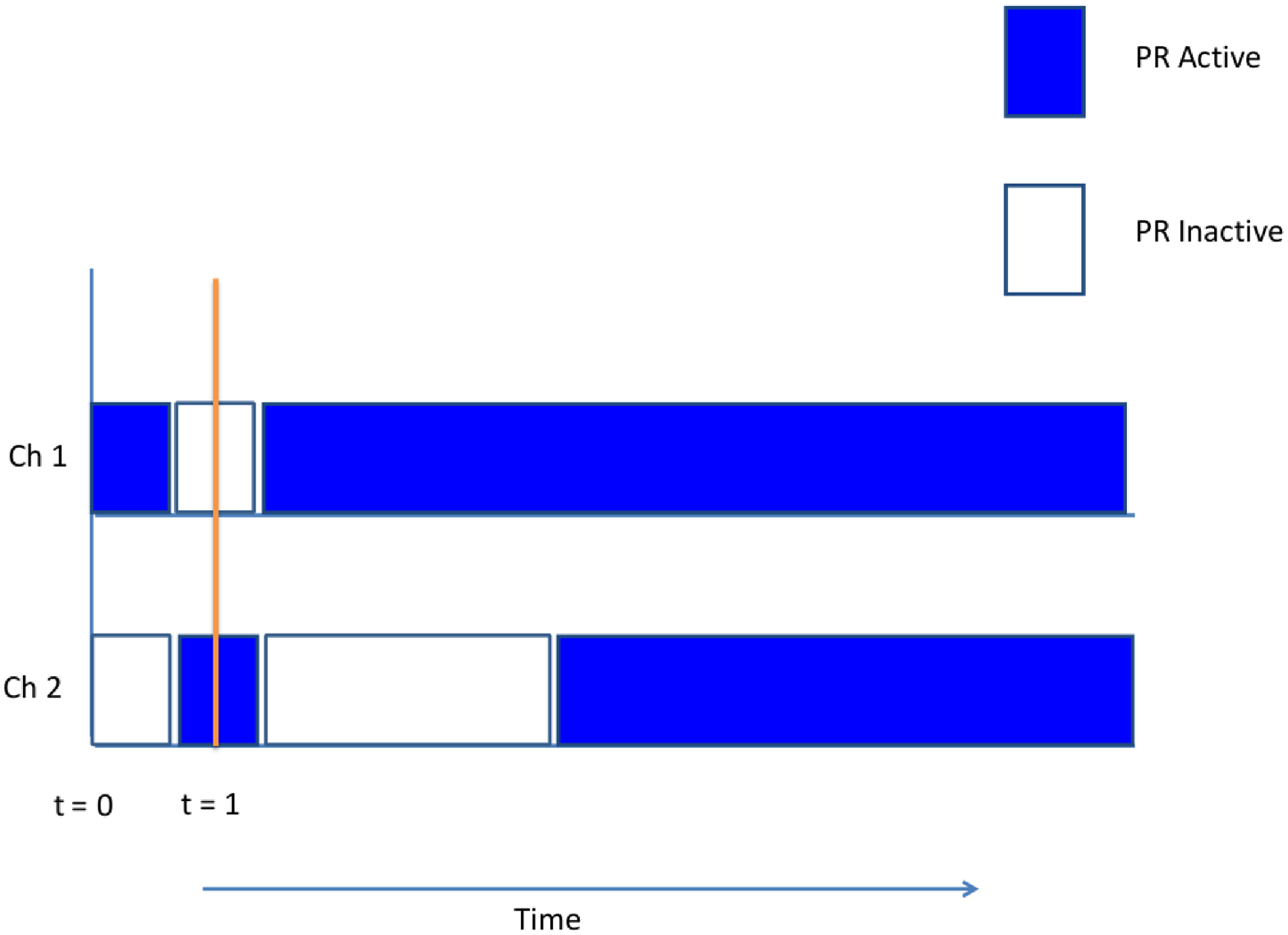}
}\hspace{-0.5cm}
\par\end{centering}

\begin{centering}

\subfigure[]
{
\label{fig:wrong-selection-long}
	\epsfxsize= 8cm
	  \leavevmode\epsfbox{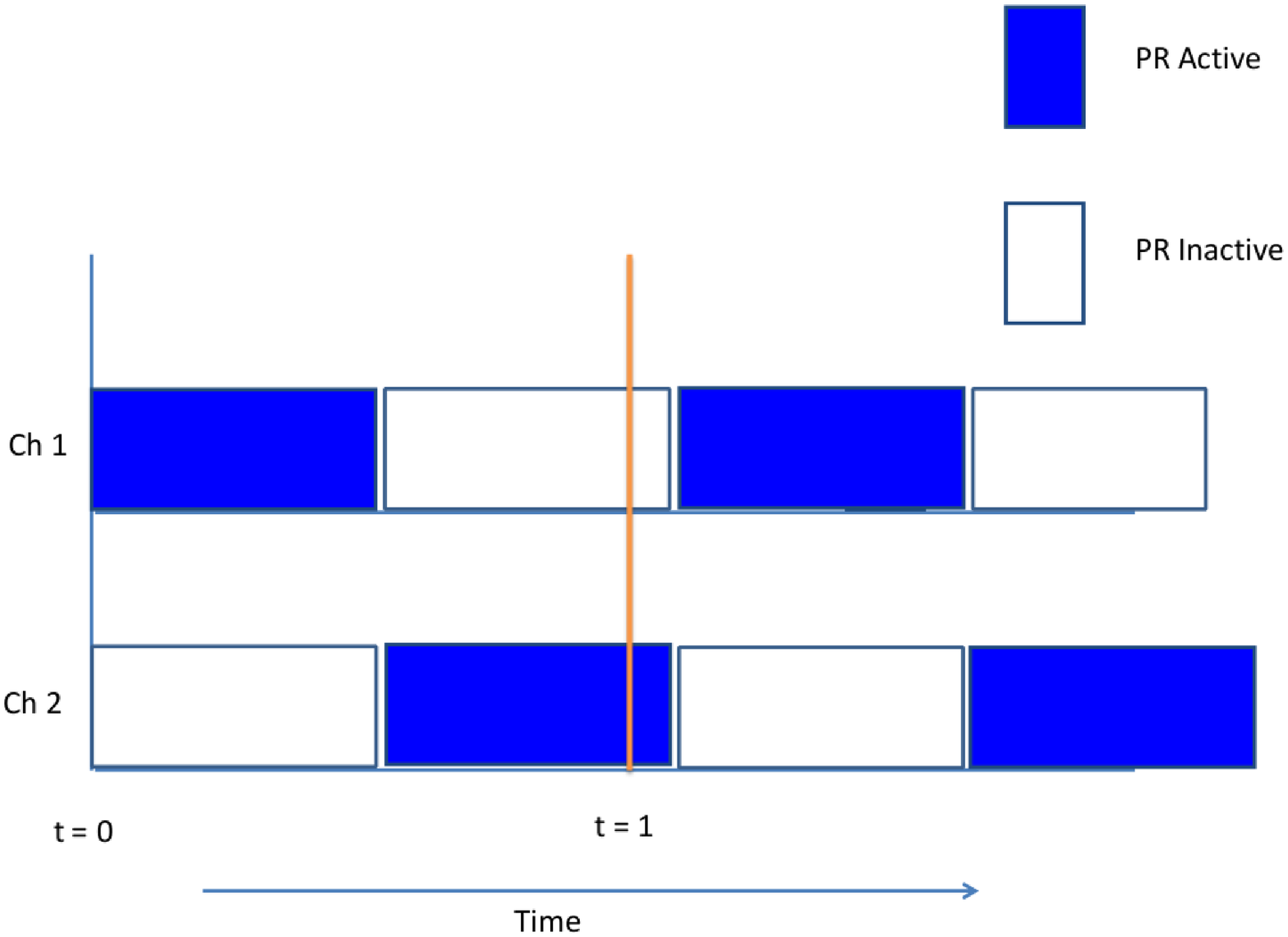}
}\hspace{-0.5cm}
\vspace{-0.3cm}
\subfigure[]
{
\label{fig:wrong-selection-inter}
	\epsfxsize= 8cm
	  \leavevmode\epsfbox{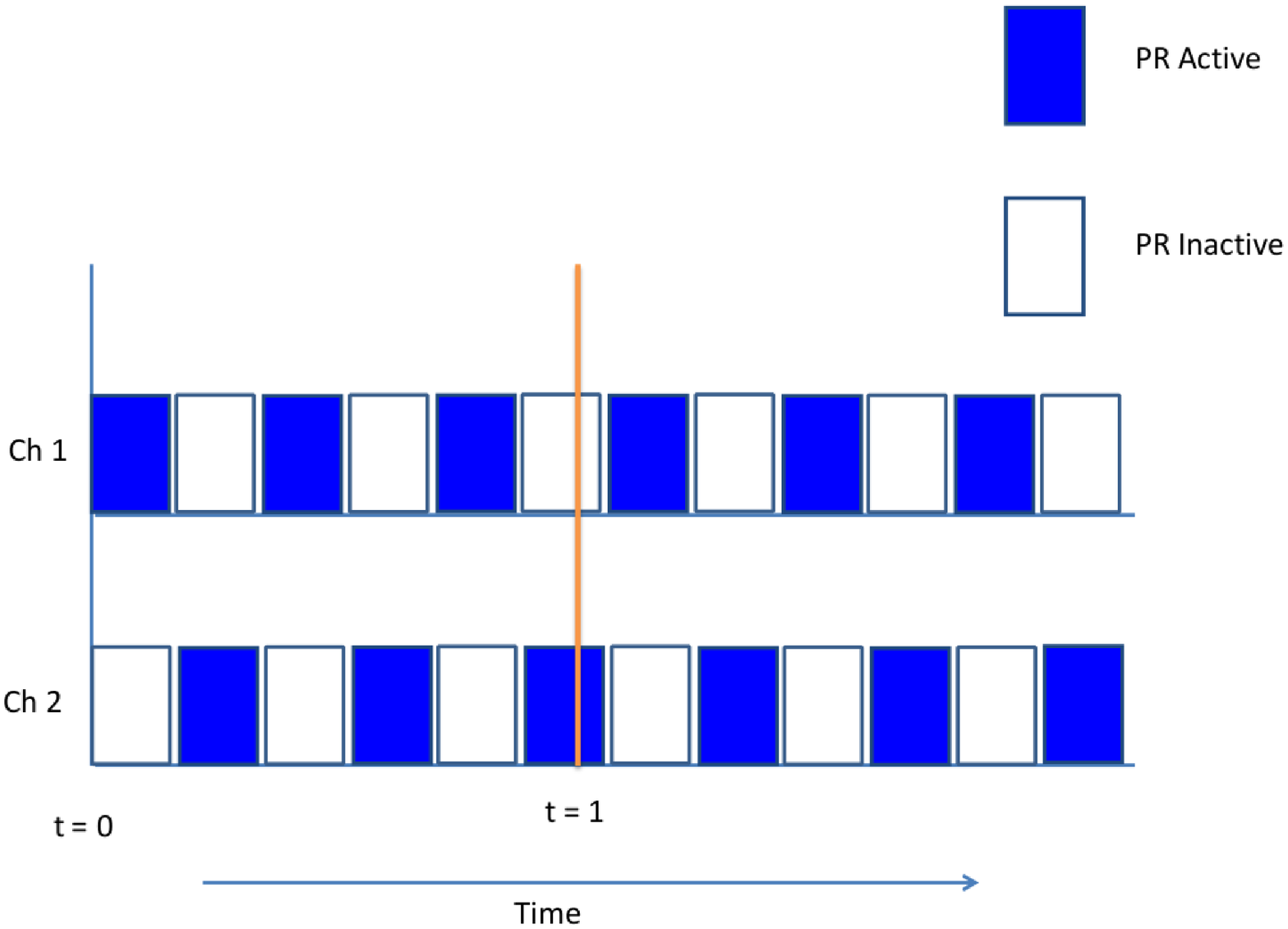}
}\hspace{-0.5cm}
\par\end{centering}

\centering{}
\subfigure[]
{
\label{fig:wrong-selection-4}
	\epsfxsize= 8cm
	  \leavevmode\epsfbox{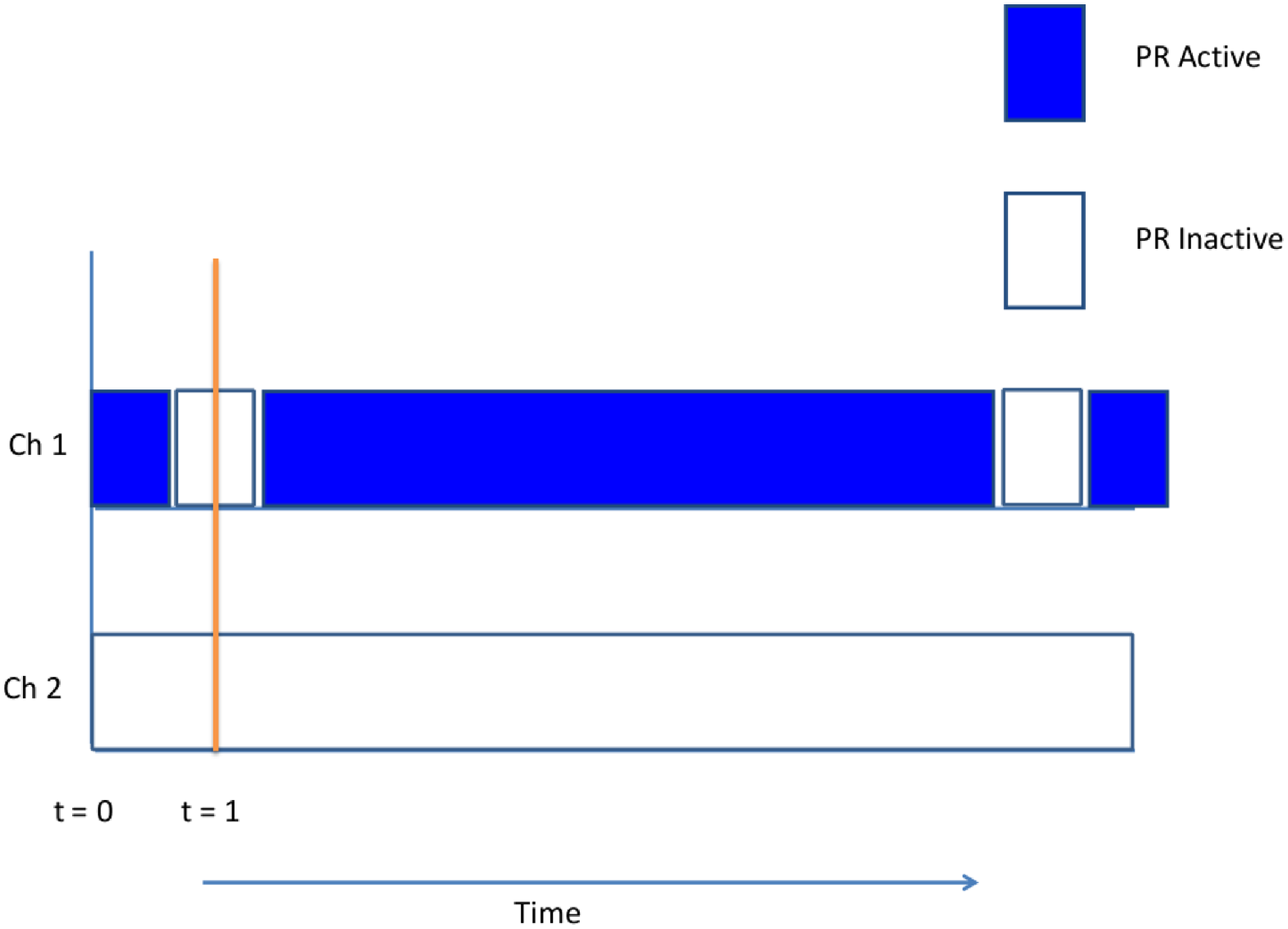}
}\hspace{0cm}
\vspace{-0.3cm}
\centering{}{\protect\caption{Instantaneous Channel Selection for CRSN nodes considering multiple PR activities (a). Wrong channel selection of Ch2 as Ch1 has low PR activity and Ch2 has high PR activity. (b). Wrong channel selection of Ch1 as Ch1 has high PR activity and Ch2 has low PR activity. (c). Wrong channel selection of Ch1 as both channels have long PR activity and idle slot of Ch1 is about to finish and Ch2 is more suitable for selection. (d). Wrong channel selection of Ch1 as both channels have intermittent PR activity and Ch1 is about to get busy and Ch2 will be available after a short while. (e). Wrong channel selection of Ch1 as it has high PR activity and Ch2 is all available hence suitable for CRSN communication.   \label{fig:wrong-channels-selections}}}
\end{center}
\end{figure*}

\subsection{Importance of Remaining Idle Time Aware Channel Selection Decision}

We assume that we have a multimedia CRSN node and when it has some data to send, it needs to perform CB. There are multiple channels and CRSN needs to select the suitable channel for communication. Now, it is required that CRSN node must not create interference for the PR node. The channel selection will be done using the three step process known as \textit{cognitive cycle} \cite{haykin2005cognitive, mitola1999cognitive}. In determining the vacant channel, the CRSN node can either follow an instantaneous channel selection method or remaining idle time information can be used.

Lets analyze that if CRSN node uses instantaneous channel selection method, will it be an effective way or not? Instantaneous channel selection is a method which searches for a vacant channel at any instant of time when a CRSN node wants to acquire a channel. It has been demonstrated in Fig. \ref{fig:wrong-channels-selections} that if a node only uses current state information to select a channel, it is highly possible that it will cause interference with PR node. The node will have to face frequent channel switching and ultimately will fail to deliver the required performance. However, if a node uses prediction and current channel state to perform channel selection, it is expected that it will create less interference with PR nodes and have to perform less channel switching. The channel which is vacant at given time gets selected and assigned to CRSN node. This method does not uses any prior information of traffic over that channel so it is un-aware of probability that when PR node can re-appear on the same channel. As illustrated in Fig. \ref{fig:wrong-selection}, when CRSN node performs channel sensing using instantaneous method at time t=1, the channel 1 is busy where as channel 2 seems suitable as it is free at that instant of time so channel 1 gets selected. However, just after that instant, channel 1 becomes free and channel 2 gets occupied by PR node and CRSN node will have to stop its communication, leave that channel and will again need to perform channel sensing. Similarly, in Fig. \ref{fig:wrong-selection-3}, channel 1 gets selected as it is vacant at that instant of time but it is not suitable for CRSN communication. 

The same channel selections have been illustrated in Fig. \ref{fig:wrong-selection-long} and \ref{fig:wrong-selection-inter} where CRSN node mistakenly selects those channel which are not suitable for its communication. A unique channel selection scenario is depicted in Fig. \ref{fig:wrong-selection-4}, when a node performs channel sensing at t=1 and both channels were available at that time. The CRSN node selects the first available channel i.e channel 1, but it gets occupied by PR node very soon however, channel 2 was vacant for a long duration of time but was not selected. It is due to the reason that instantaneous channel selection does not considers the PR activity model and performs channel selection based on current channel sensing result only.

\begin{algorithm}[ht]
  \caption{\footnotesize{Remaining Idle Time Aware Channel Bonding Algorithm for CRSNs (RITCB-IP)}}\label{Alg-3}
\scriptsize
1: $CH$: Total number of channels\\
2: $CBsize$: Size for channel bonding\\
3: $SCH$: Selected channels\\
4: $CCH$: Contiguous channels\\ 
5: $RIT_i$: Remaining idle time of $i^{th}$ channel\\
6: $SRIT$: Smallest remaining idle time\\
7: $RITCH-IP3$: The array of three contiguous channels with RIT\\
8: $RITCH-IP2$: The array of two contiguous channels with RIT\\
9: \textbf{Start}\\
10: Perform spectrum sensing to find total number of channels $CH$\\
11: Input $CH$, $CBsize$\\
12: Select all the possible channel combinations of $CBsize$ from $CH$\\
13:	selected\_channels $\leftarrow$ add the value of possible combinations\\
14: $SCH$ $\leftarrow$ arrange the selected channels in descending order\\ 
15:	goto algorithm \ref{Alg-4} $\leftarrow$ to find set of contiguous channels\\
16:	\tab\textbf{for} ${i=0, i < CCH, i++}$ \\
17:		\tab\tab $CONT[CCH]$ $\leftarrow$ Arrange the contiguous channel pairs in descending order\\
18: \tab \tab Perform detailed channel sensing on contiguous channels to find PR activity on $CONT[CCH]$\\
19:		\tab\tab $CONT[CCH3]$ $\leftarrow$ Insert the sets of three contiguous pairs from $CONT[CCH]$\\
20:		\tab\tab $CONT[CCH2]$ $\leftarrow$ Insert the sets of two contiguous pairs from $CONT[CCH]$\\
21:	\tab\textbf{end for}	\\	
22:	 \tab\textbf{for} ${j=0, j < CONT[CCH3], j++}$ \\
23:		\tab\tab For All $j$ Compute $RIT$\\ 
24:		\tab\tab $SRIT[CCH3]$ $\leftarrow$ $min$ [$RIT(CONT[CCH3])$]\\
25:	 \tab\textbf{end for} \\
26: \tab $RITCH-IP3$ $\leftarrow$ Select the channels from $SRIT[CCH3]$ with longest $RIT$\\
27:	 \tab\textbf{for} ${k=0, k < CONT[CCH2], k++}$ \\
28:		\tab\tab For All $k$ Compute $RIT$\\ 
29:		\tab\tab $SRIT[CCH2]$ $\leftarrow$ $min$ [$RIT(CONT[CCH2])$]\\
30:	 \tab\textbf{end for} \\
31: \tab $RITCH-IP2$ $\leftarrow$ Select the channels from $SRIT[CCH2]$ with longest $RIT$\\
32:	\tab\textbf{if} $RITCH-IP3$ $>$ $RITCH-IP2$ \textbf{then}\\	
33:		\tab\tab Make a bond using three contiguous $RITCH-IP3$\\
34:		\tab\tab\tab\textbf{if}	PR node active on selected bond of $RITCH-IP3$ \textbf{then}\\
35:		\tab\tab \tab \tab break the bond, drop the packet and release the channels for PR node\\
36:		\tab\tab \tab \textbf{else if} PR node is not active on selected bond of $RITCH-IP3$ \textbf{then}\\
37:		\tab\tab\tab\tab Transmit data\\
38:		\tab\tab\tab\tab Break the bond and release the channels\\
39:	\tab\textbf{else if} $RITCH-IP3$ $==$ $RITCH-IP2$ \textbf{then}\\
40:		\tab\tab Make a bond using three contiguous $RITCH-IP3$\\
41:		\tab\tab\tab\textbf{if}	PR node active on selected bond of $RITCH-IP3$ \textbf{then}\\
42:		\tab\tab \tab \tab break the bond, drop the packet and release the channels for PR node\\
43:		\tab \tab\tab \textbf{else if} PR node is not active on selected bond of $RITCH-IP3$ \textbf{then}\\
44:		\tab\tab\tab Transmit data\\
45:		\tab\tab\tab Break the bond and release the channels\\
46:	\tab\textbf{else} $RITCH-IP3$ $<$ $RITCH-IP2$ \textbf{then}\\
47:		\tab\tab Make a bond using two contiguous $RITCH-IP2$\\
48:		\tab\tab\tab\textbf{if}	PR node active on selected bond of $RITCH-IP3$ \textbf{then}\\
49:		\tab\tab \tab \tab break the bond, drop the packet and release the channels for PR node\\
50:		\tab \tab\tab \textbf{else if} PR node is not active on selected bond of $RITCH-IP3$ \textbf{then}\\
51:		\tab\tab\tab Transmit data\\
52:		\tab\tab\tab Break the bond and release the channels\\
53: \textbf{Stop}\\

\end{algorithm}

\begin{algorithm}[t]
\caption{\footnotesize{Algorithm to find Contiguous Channels for RITCB-IP}}\label{Alg-4}
\scriptsize
1: \textbf{Start}\\
2: Input $SCH$\\
3: For each $SCH$ pair, get the channel ID's\\
4: \tab \textbf{if} all three channels in $SCH$ are consecutive \textbf{then}\\
5:		\tab \tab $CCH$ $\leftarrow$ all selected channels are contiguous\\
6:		\tab \tab shortlist the ID's of three contiguous channels\\
7:		\tab \textbf{else if} any two channels in $SCH$ are consecutive \textbf{then}\\
8:		\tab \tab $CCH$ $\leftarrow$ two selected channels are contiguous\\
9:		\tab \tab shortlist the ID's of two contiguous channels\\
10: \textbf{Stop}\\
11: goto Algortithm \ref{Alg-3}\\
\end{algorithm}

\subsection{Working of Remaining Idle Time Aware Intelligent Channel Bonding Schemes RITCB and RITCB-IP}

In this section, we discuss the working of our proposed algorithms. Our proposed remaining idle time aware channel bonding schemes RITCB and RITCB-IP perform channel selection based on PR activity over the channels. As shown in algorithm. \ref{Alg}, first RITCB takes number of channels and CB size as input. According to the given parameters, the algorithm selects the channels and arranges them. The algorithm \ref{Alg-2} finds out that the selected channels have three contiguous channels or two contiguous channels or they are not contiguous. \ans{All the pairs of three contiguous and two contiguous channels are arranged according to size in descending order in $CONT[CCH]$. The $CONT[CCH]$ is further classified into three ans two contiguous channels pairs as $CONT[CCH3]$ and $CONT[CCH2]$ respectively.} Upon finding the contiguous channels, it performs detailed channel sensing, detects the PR activity over the channels and then estimates that the channels will be available for how much duration of time. The selection of channels is done on the basis of remaining idle time of PR node. The longer the PR node will remain idle over the channel the more suitable is the channel for CR node. 

After the selection of channels, the RIT of all contiguous sets having three channels is calculated. When RIT values have been achieved then, the minimum RIT value of each contiguous set is calculated and stored in SRIT. It reduces the possibility of creating harmful interference with PR nodes. When SRIT of all contiguous sets is achieved then the maximum RIT of from SRIT is selected and stored in RTICH3. The same criteria is followed for contiguous sets of two channels and the maximum RIT is stored in RITCH2. Now, we compare the values of RITCH3 and RITCH2 to select the size of bond. Our target is to select the larger bond size having maximum RIT so bond of two channels is selected only if RITCH2 $>$ RITCH3. The contiguous pair having the longest remaining idle time is the output of our proposed algorithm. In this way, our proposed scheme selects the suitable channels for CRSN nodes and these channels can be combined to perform channel bonding.


In presence of PR activities, the number of RIT based channels which get selected varies. We also increase the number of channels to check the behavior of algorithm in presence of larger channels set. When there is low PR activity, the RITCB finds the set of 3 contiguous channels with longest RIT when the number of channels is increased. In contrast, the set of 2 contiguous channels with longest RIT get selected when there is high PR activity on the network. It is due to the reason that channels are mostly busy and it is impossible to get 3 contiguous channels with longest RIT with high PR activity. 

\begin{figure}
  \begin{center}
    {
      \epsfxsize= 8cm
	  \leavevmode\epsfbox{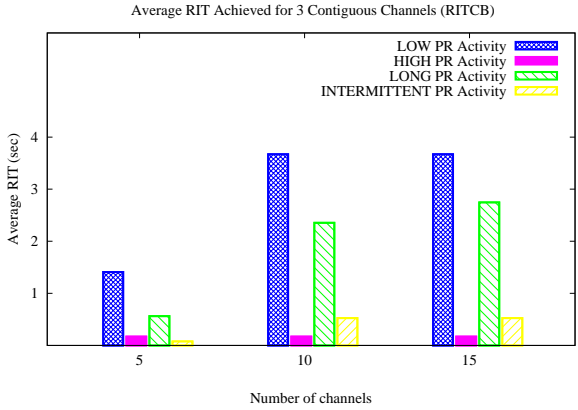}
    }\hspace{0cm}
\centering{}\caption{Average RIT based Time for all PR activities}\vspace{-0.7cm}
\label{fig:RIT-avg}
\end{center}
\end{figure}

We can also estimate the average remaining idle time of selected contiguous pair. It is elaborated in Fig. \ref{fig:RIT-avg} that as number of channels in the network increase, the possibility of finding longer RIT increases. When low PR activity is present in the network, we get longer RIT as compared to high PR activity which confirms the effectiveness of our proposed scheme. Long PR activity also provide considerable amount of RIT as there are long OFF periods which can be utilized for CB purpose.

The RITCB-IP works differently in case of HIR avoidance as shown in algorithm \ref{Alg-3}. When the contiguous channels based on RIT get selected, RITCB-IP checks that if PR node is active on any channel of the bonded pair. In case of PR presence, the RITCB-IP stops the CRSN node communication by dropping the packet and releasing the channels for PR node. In this way, RITCB-IP effectively prevents harmful interference with PR nodes.

\section{Performance Evaluation}

In this section, we discuss performance evaluation of our proposed algorithms and compare them with PRACB, SWA, KNOWS and AGILE.

\subsection{Simulation Results}

We have done performance analysis of our proposed schemes RITCB and RITCB-IP by comparing them with PRACB \cite{bukhari2016pracb}, SWA \cite{Chandra2008}, KNOWS \cite{Yuan2007} and AGILE \cite{Keranidis2014}. We have decided to compare our approaches with SWA,  KNOWS, and AGILE, as these schemes are focusing on channel bonding. We did not compare our approaches with RIT based single channel selection schemes proposed for CRNs (c.f. Table. \ref{tab:RIT-schemes})  because these single channel schemes are generally not well suited for channel bonding (c.f. \cite{bukhari2016pracb}). In PRACB \cite{bukhari2016pracb}, the channels get selected randomly and only those channels are used for CB which are free from PR activity at that time. The channels which have active PR nodes are dropped for CR utilization. SWA \cite{Chandra2008} applies the concept of CB by varying the channel width without considering any PR activity hence it causes interference with PR nodes. KNOWS \cite{Yuan2007} utilizes the white spaces in TV band for channel width adaptation. AGILE \cite{Keranidis2014} tunes the node central frequency and bandwidth to the under utilized WLAN spectrum to use it for unlicensed operations. \ans{Earlier, we have incorporated the CRSN capabilities in NS-2 \cite{bukhari2016ns}}. We have now thoroughly tested the proposed scheme by simulating it in NS-2. The CRSN nodes first find the contiguous channels and then check the PR activity over the selected channels. In addition, they select channels on the basis of remaining idle time as discussed in previous section. The simulation parameters have been mentioned in Table. \ref{Tab:Simulation-Parameters}. \ans{There are two CRSN nodes which establish single hop communication in presence of four types of PR activities. In this work, we show that how intelligent channel selection gives the suitable channels for channel bonding which will be idle for longest possible duration in a single hop scenario. The implementation of RITCB and RITCB-IP in multi hop scenario can be considered as future work.}

\begin{table}[]
\begin{center}
\resizebox{\columnwidth}{!}{%
\begin{tabular}{|c|c|}
\hline 
Parameters & Values\tabularnewline
\hline 
\hline 
Simulator Name & NS-2\tabularnewline
\hline
Patch used for simulating CRSN & CRCN patch \tabularnewline
\hline
Operating frequency band & 2.4 GHz \tabularnewline
\hline
Number of nodes & 02\tabularnewline
\hline 
Number of channels & 3 - 15\tabularnewline
\hline 
Size of simulation area & 1000{*}1000\tabularnewline
\hline 
Radio propagation model & Two-ray ground model\tabularnewline
\hline 
Network interface type & Wireless PHY\tabularnewline
\hline 
Routing protocol & AODV\tabularnewline
\hline 
MAC type & Mac/Maccon\tabularnewline
\hline 
PR activity & ON / OFF with exponential distribution\tabularnewline
\hline 
PR activity regimes & Low, High, Long and Intermittent\tabularnewline
\hline 
Sensor initial energy ($E_{in}$) & 1J \tabularnewline
\hline
Transceiver energy consumption ($E_{tr}$) &  50nj \cite{Shah2015}\tabularnewline
\hline
Simulation time & 10000 sec\tabularnewline
\hline 
No. of packets sent & 10000\tabularnewline
\hline 
Packet size & 44 bytes\tabularnewline
\hline 
Size of bonded channels & 3 bonded channels / 2 bonded channels\tabularnewline
\hline 

\end{tabular}%
}

\centering{}\protect\caption{Simulation Parameters \label{Tab:Simulation-Parameters}}

\end{center}
\end{table}

\begin{figure*}[htbp]
  \begin{center}
    \subfigure[]
    {
      \label{RIT-dr-low}
      \epsfxsize= 7cm
	  \leavevmode\epsfbox{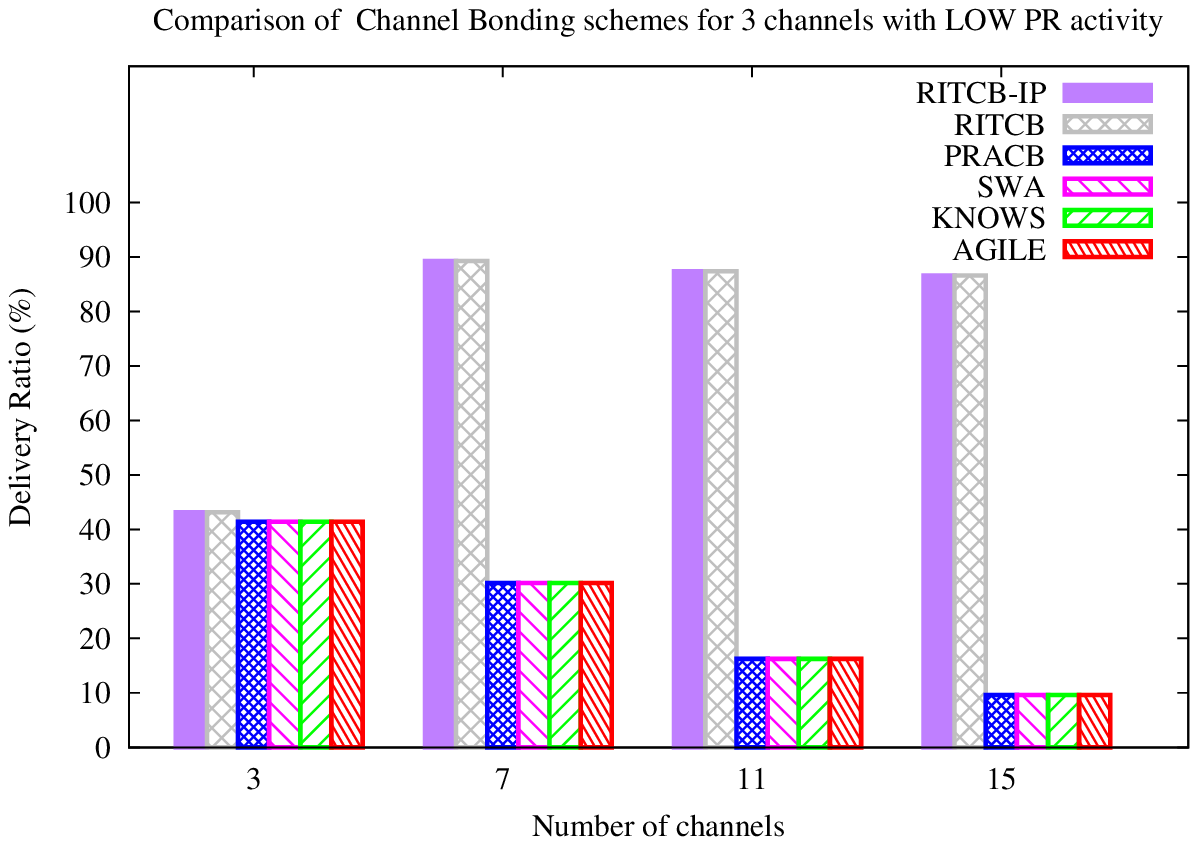}
    }\hspace{0cm}
    \subfigure[]
    {
      \label{RIT-dr-high}
      \epsfxsize= 7cm
	  \leavevmode\epsfbox{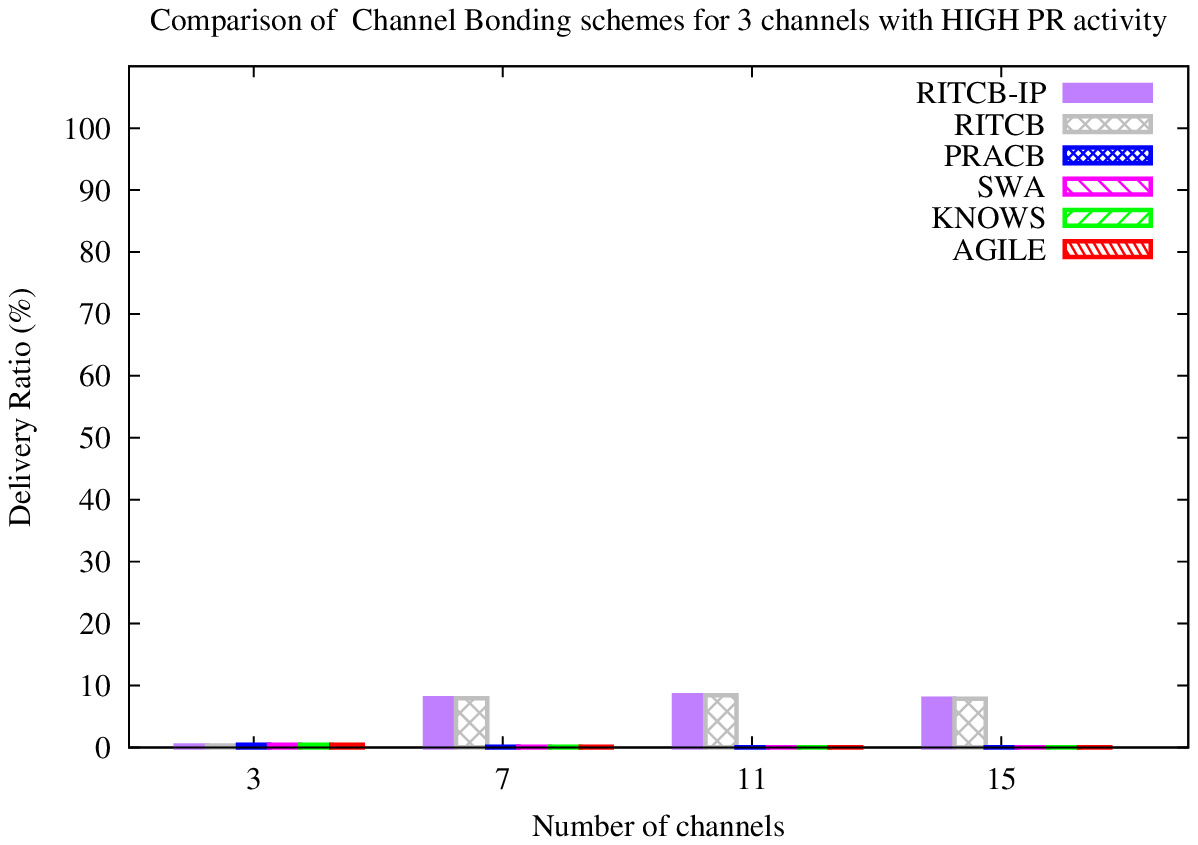}
    }\hspace{0cm}
    \subfigure[]
    {
      \label{RIT-dr-long}
      \epsfxsize= 7cm
	  \leavevmode\epsfbox{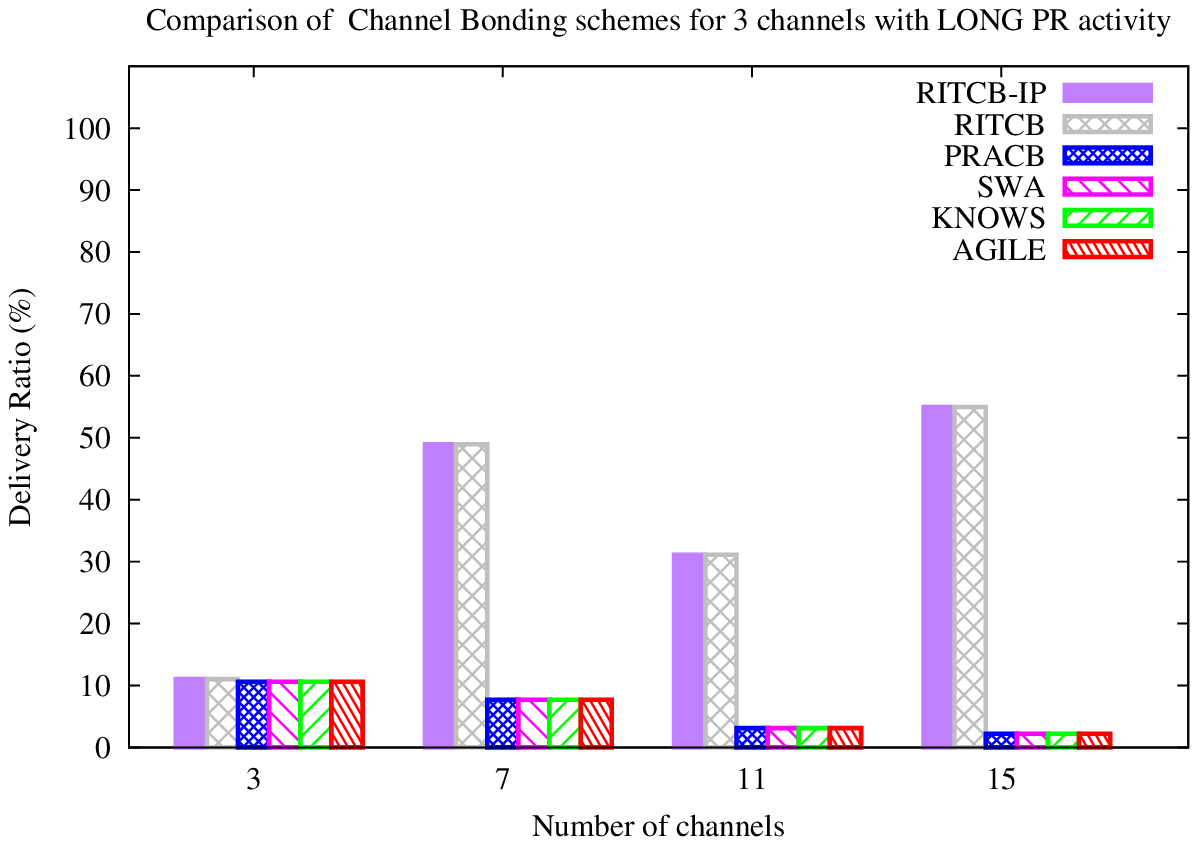}
    }\hspace{0cm}
  \subfigure[]
    {
      \label{RIT-dr-inter}
      \epsfxsize= 7cm
      \leavevmode\epsfbox{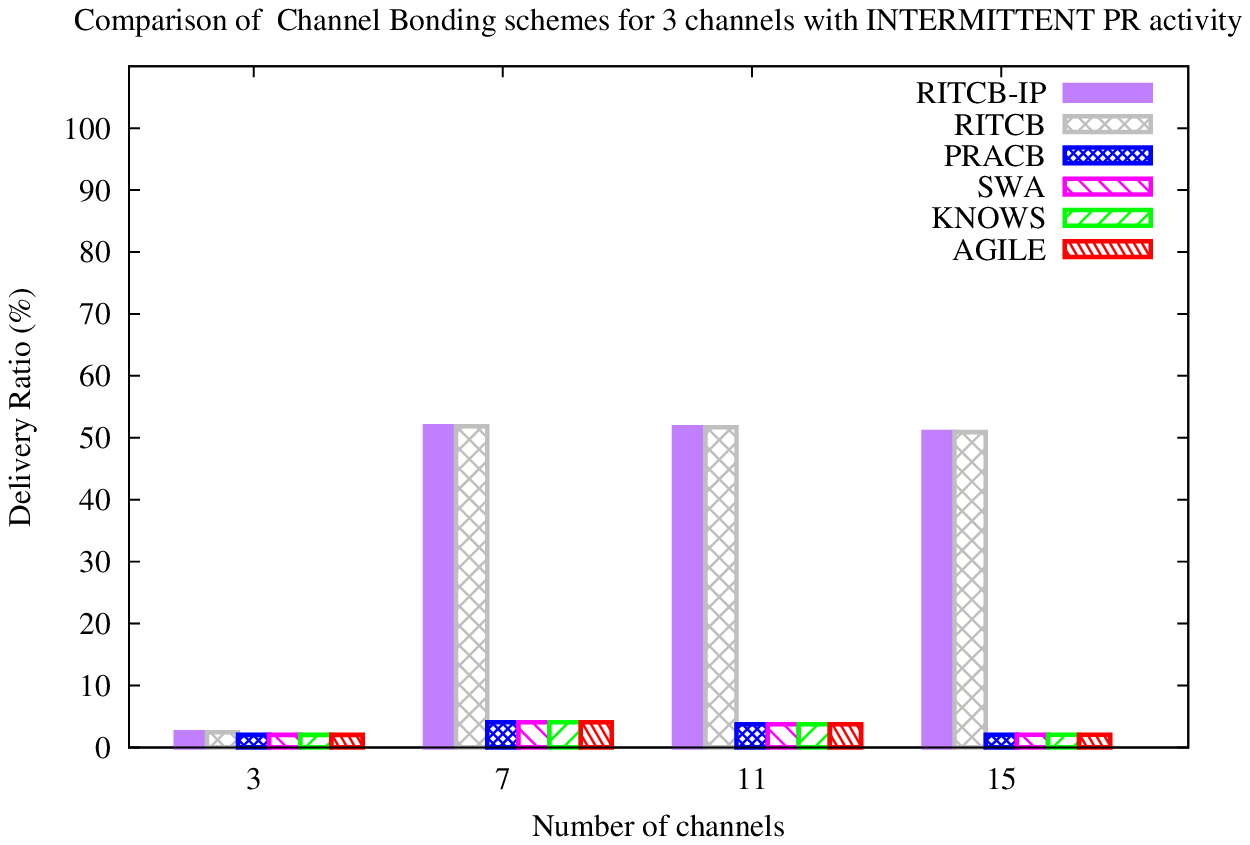}
    }\hspace{0cm}
   \vspace{-0.3cm} \caption{Overview of Delivery Ratio for RIT based contiguous channels with: (a) Low PR Activity. (b) High PR Activity. (c) Long PR Activity. (d) Intermittent PR Activity.}\vspace{-0.7cm}
    \label{Overview-of-dr-all-PR}
  \end{center}
\end{figure*}

\subsubsection{Impact of intelligent channel selection on PR activity model}

As we have discussed earlier, PR activity model plays vital role in the performance of CRSN nodes. The availability of channels depends upon the type of PR activity present in the network. The selection of available channels is a critical problem which impacts the throughput of CRSN nodes. The intelligent channel selection provides an opportunity to CRSN nodes that they can select those channels which are available for longer duration of time as compared to other channels. This intelligent channel selection improves the channel utilization and other parameters as discussed later in this section.

\subsection{Comparison Analysis}

Now we discuss the comparison analysis of our proposed schemes RITCB and RITCB-IP by comparing with other four schemes based on following performance parameters.

\begin{table}[]
\begin{center}
\footnotesize

\begin{tabular}{|c|c|c|}
\hline 
Sr. No. & Scheme & Complexity\tabularnewline
\hline 
\hline 
1. & RITCB-IP & O($n$)\tabularnewline
\hline
2. & RITCB & O($n$) \tabularnewline
\hline
3. & PRACB & O($n^2$) \tabularnewline
\hline
4. & SWA & O($n$) \tabularnewline
\hline 
5. & KNOWS & O($n$) \tabularnewline
\hline 
6. & AGILE & O($n$) \tabularnewline
\hline 

\end{tabular}%

\centering{}\protect\caption{Complexity Analysis \label{Tab:Complexity-Analysis}}

\end{center}
\end{table}

\subsubsection{Complexity Analysis}

\ans{Time complexity is important for any algorithm to estimate the maximum time required to execute. We have used Big O notation to describe the performance of our proposed schemes. Big O notation specifically provides the worst-case scenario for any scheme in terms of time or required memory. As mentioned in Table. \ref{Tab:Complexity-Analysis}, our proposed schemes RITCB-IP and RITCB have the complexity as O$(n)$ which shows the linearity of our schemes as compared to PRACB which has the complexity as O$(n^2)$. The other schemes such as SWA, KNOWS and AGILE also has linear complexities i.e. O$(n)$ but our proposed schemes outperform in other parameters such as delivery ratio, harmful interference, node life and channel switching. }

\subsubsection{Delivery Ratio}

Delivery Ratio can be defined as the successful number of packets received from total sent packets. Suppose a CRSN node is utilizing a channel to send its data to a sink node then how many packets it successfully transmits out of all packets sent is called as delivery ratio. 

In our simulation scenario, the CRSN node selects RIT based pair of contiguous channels to make a bond and transmits the data using RITCB and RITCB-IP. We count the total number of packets transmitted and number of packets successfully received at receiver. The results in Fig. \ref{Overview-of-dr-all-PR} show that RITCB and RITCB-IP provide significant better delivery ratio than PRACB, SWA, KNOWS and AGILE. The RITCB and RTICB-IP give same delivery ratio as they select the channel following the same criteria i.e. RIT based contiguous channels selection. 

The major improvement in delivery ratio can be seen in case of high PR activity i.e. Fig. \ref{RIT-dr-high} where our proposed schemes are giving visible delivery ratio when other schemes are almost zero. It is due to the intelligent channel selection of contiguous channels based on remaining idle time.

The significant performance in terms of delivery ratio can also be seen in other PR activities such as low, long and intermittent (Fig. \ref{RIT-dr-low}, \ref{RIT-dr-long}, and \ref{RIT-dr-inter} respectively).

\textit{Hence, it can be concluded that intelligent channel selection through RIT based contiguous channels can significantly improve the overall delivery ratio for all types of PR activities.}

\begin{figure*}[htbp]
  \begin{center}
    \subfigure[]
    {
      \label{RIT-hir-low}
      \epsfxsize= 7cm
	  \leavevmode\epsfbox{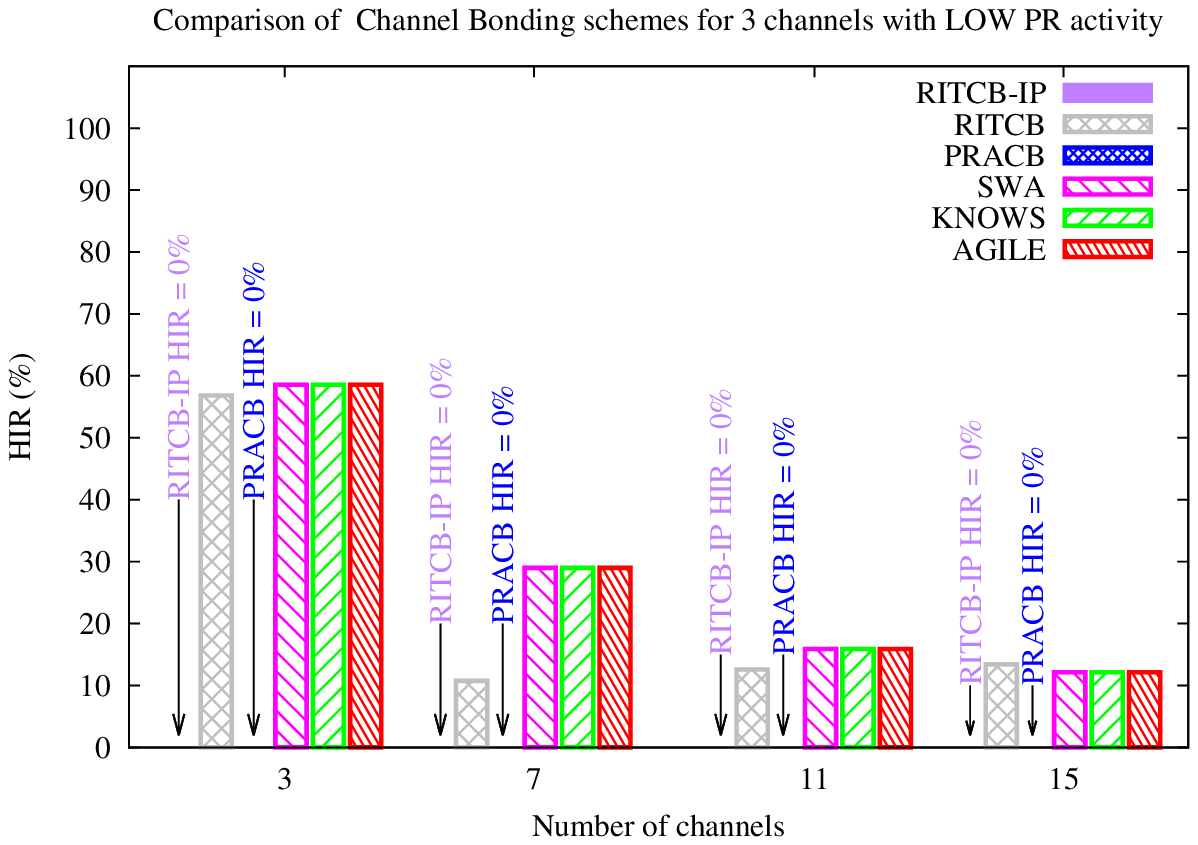}
    }\hspace{0cm}
    \subfigure[]
    {
      \label{RIT-hir-high}
      \epsfxsize= 7cm
	  \leavevmode\epsfbox{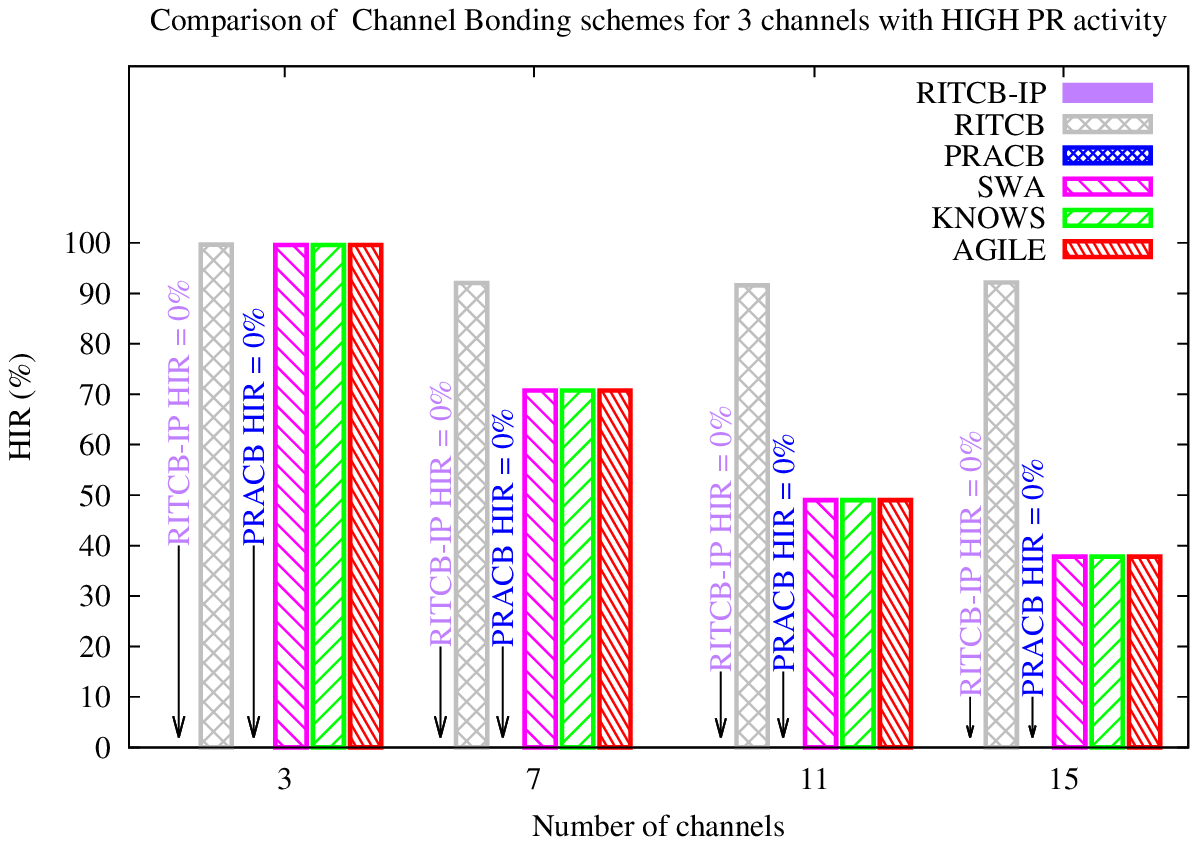}
    }\hspace{0cm}
    \subfigure[]
    {
      \label{RIT-hir-long}
      \epsfxsize= 7cm
	  \leavevmode\epsfbox{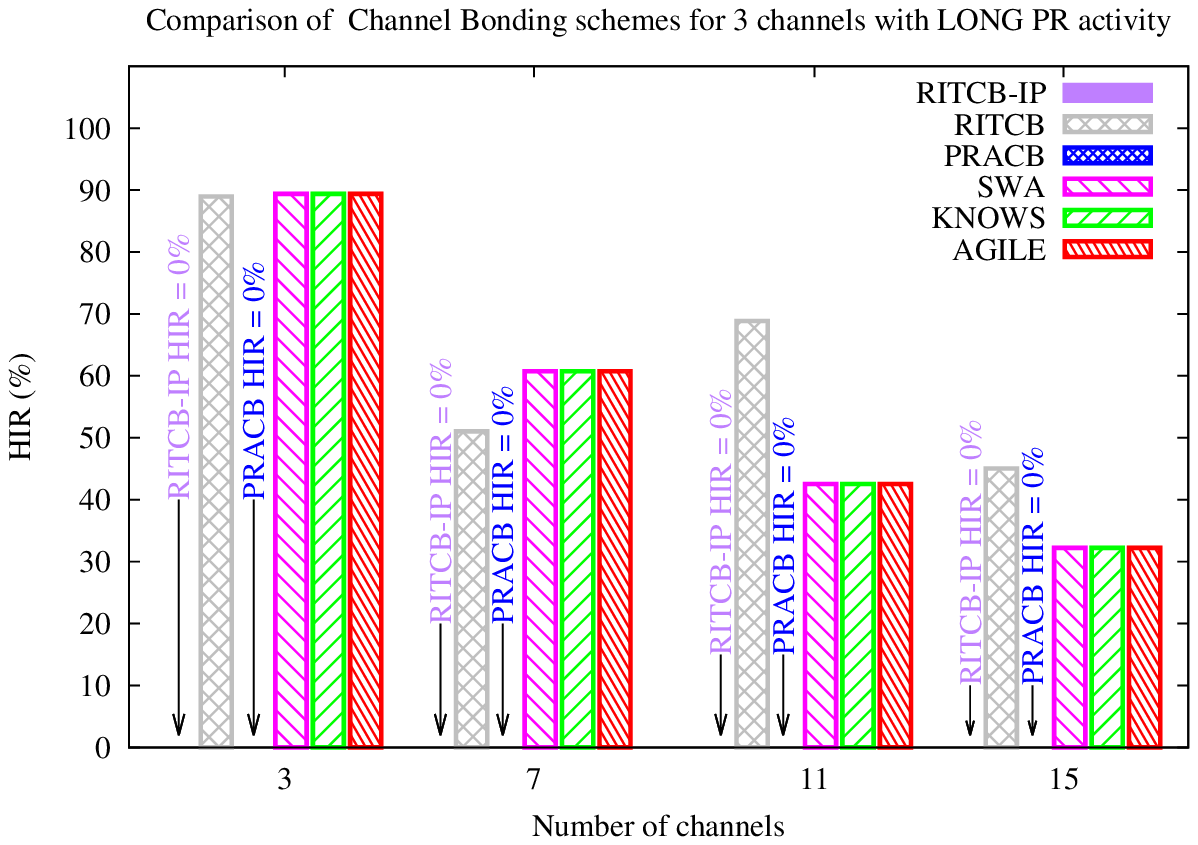}
    }\hspace{0cm}
  \subfigure[]
    {
      \label{RIT-hir-inter}
      \epsfxsize= 7cm
      \leavevmode\epsfbox{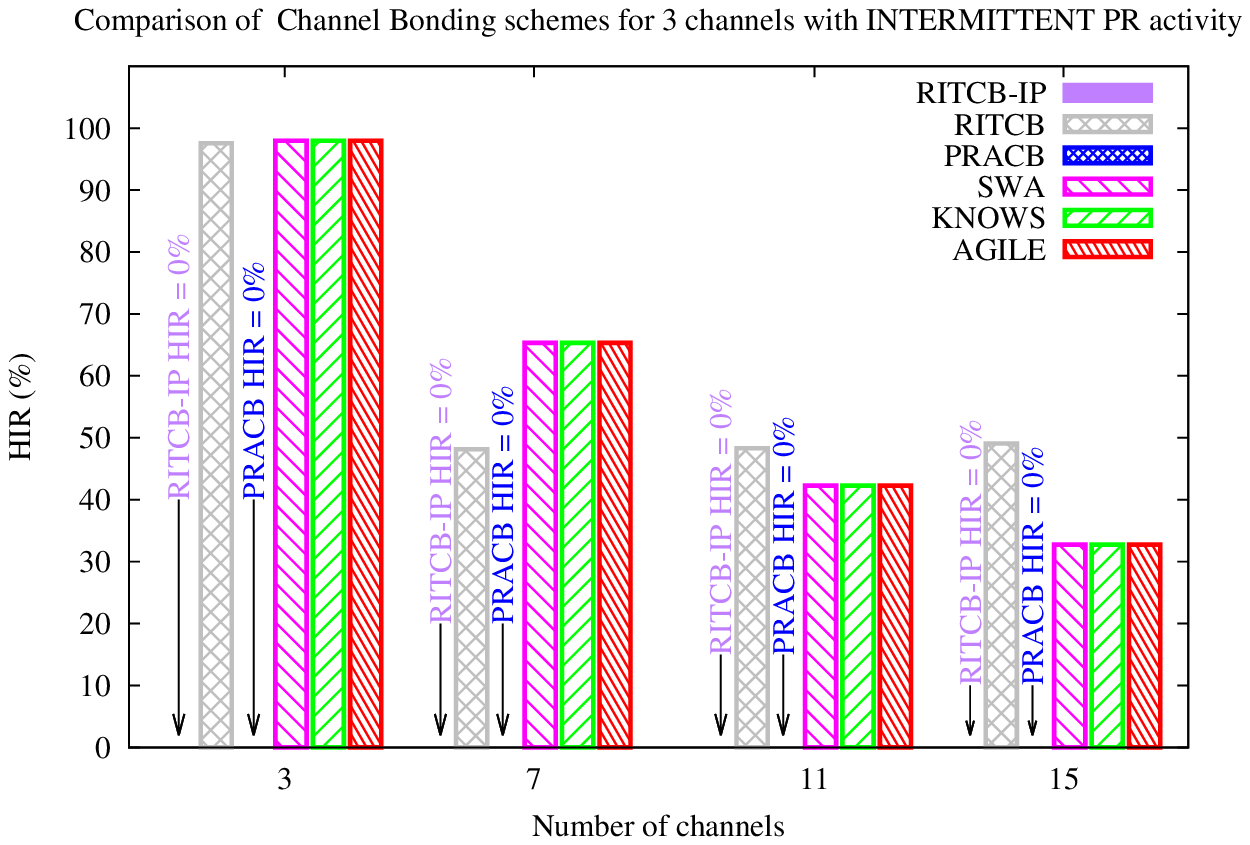}
    }\hspace{0cm}
   \vspace{-0.3cm} \caption{Overview of Harmful Interference Ratio for RIT based contiguous channels with: (a) Low PR Activity. (b) High PR Activity. (c) Long PR Activity. (d) Intermittent PR Activity.}\vspace{-0.7cm}
    \label{Overview-of-hir-all-PR}
  \end{center}
\end{figure*}

\subsubsection{Harmful Interference}

Harmful interference occurs when PR and CR nodes try to utilize channel at the same time. Suppose, a CRSN node is utilizing a channel and meanwhile a PR node appears on the same channel. The PR and CR nodes will sense interference and CR node will stop its communication. This problem is due to instantaneous channel selection by CR nodes. If CR nodes select channels on the basis of remaining idle time, then they can stop their communication before the appearance of PR nodes hence minimizing PR-CR harmful interference. 

Considering the simulation scenario, when CRSN node sends 10000 packets, we can count how many packets have been dropped due to harmful interference with PR nodes. Using RITCB, the CRSN node will cause interference because it uses the expected value of remaining idle time of channel to select that channel and when PR node appears over that channel it stops its communication (as shown in Fig. \ref{Overview-of-hir-all-PR}). However, The RITCB-IP does not causes interference with PR nodes as it performs channel sensing for every packet and sends when the channel is idle from PR activity. The HIR is increased when there is high PR activity over the network (as shown in Fig. \ref{RIT-hir-high}) as the channels are mostly occupied by PR nodes and CRSN nodes get very less chance to transmit their data.

\textit{Hence, harmful interference can be reduced by using intelligent channel selection scheme along with performing channel sensing in a periodic manner.}

\begin{figure*}[htbp]
  \begin{center}
    \subfigure[]
    {
      \label{RIT-tx-low}
      \epsfxsize= 7cm
	  \leavevmode\epsfbox{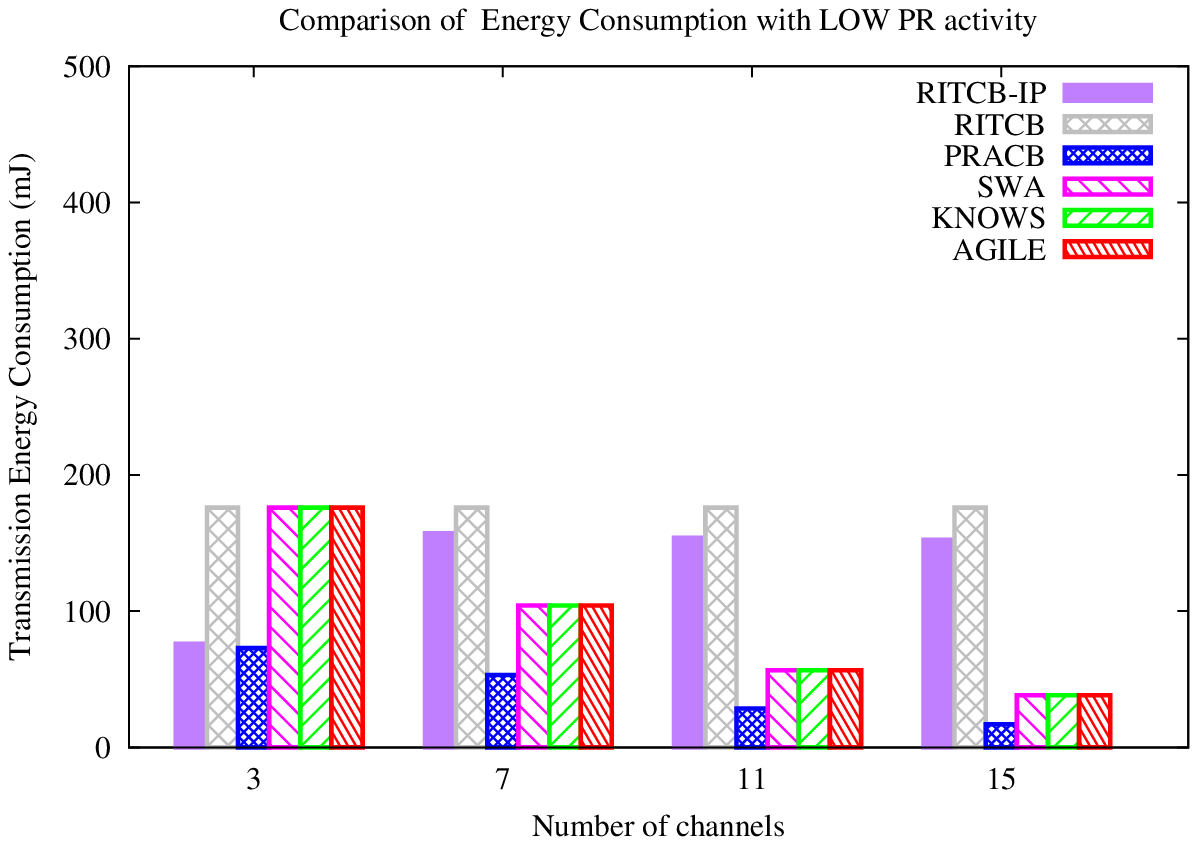}
    }\hspace{0cm}
    \subfigure[]
    {
      \label{RIT-tx-high}
      \epsfxsize= 7cm
	  \leavevmode\epsfbox{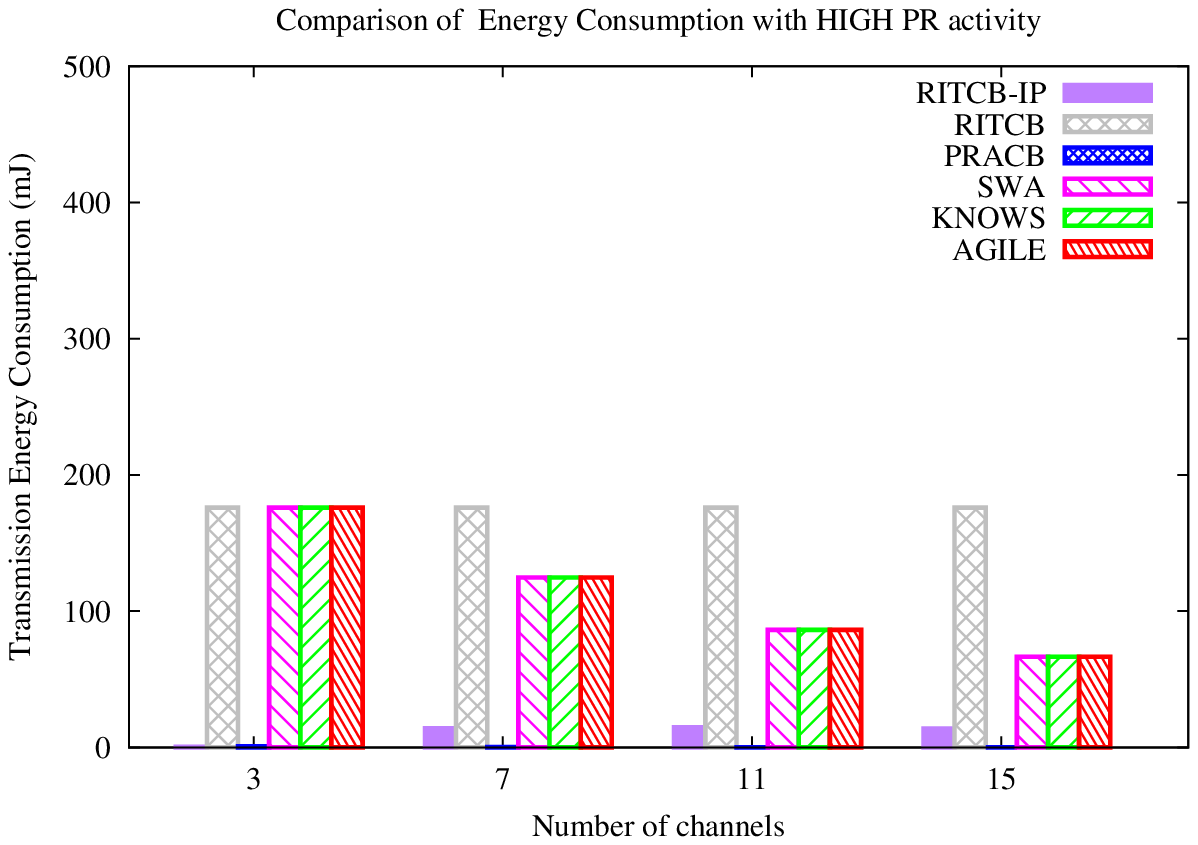}
    }\hspace{0cm}
    \subfigure[]
    {
      \label{RIT-tx-long}
      \epsfxsize= 7cm
	  \leavevmode\epsfbox{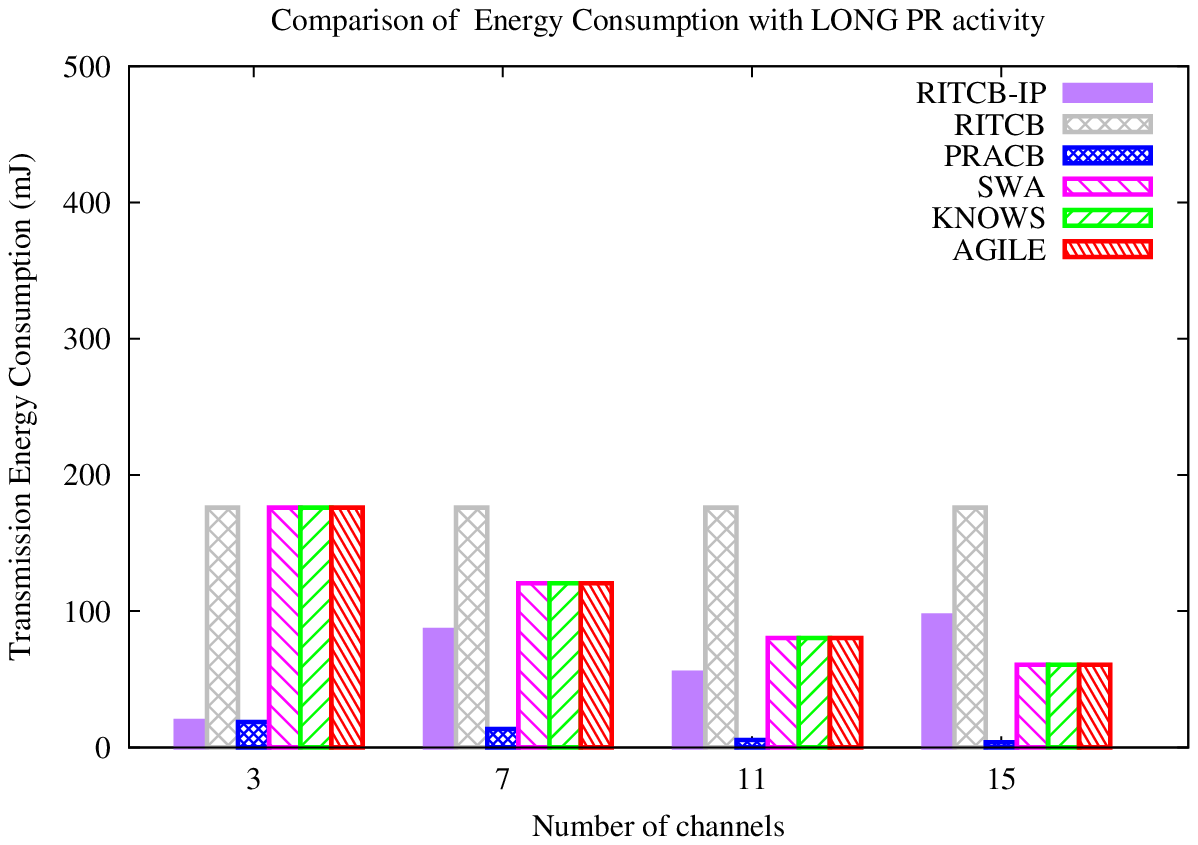}
    }\hspace{0cm}
  \subfigure[]
    {
      \label{RIT-tx-inter}
      \epsfxsize= 7cm
      \leavevmode\epsfbox{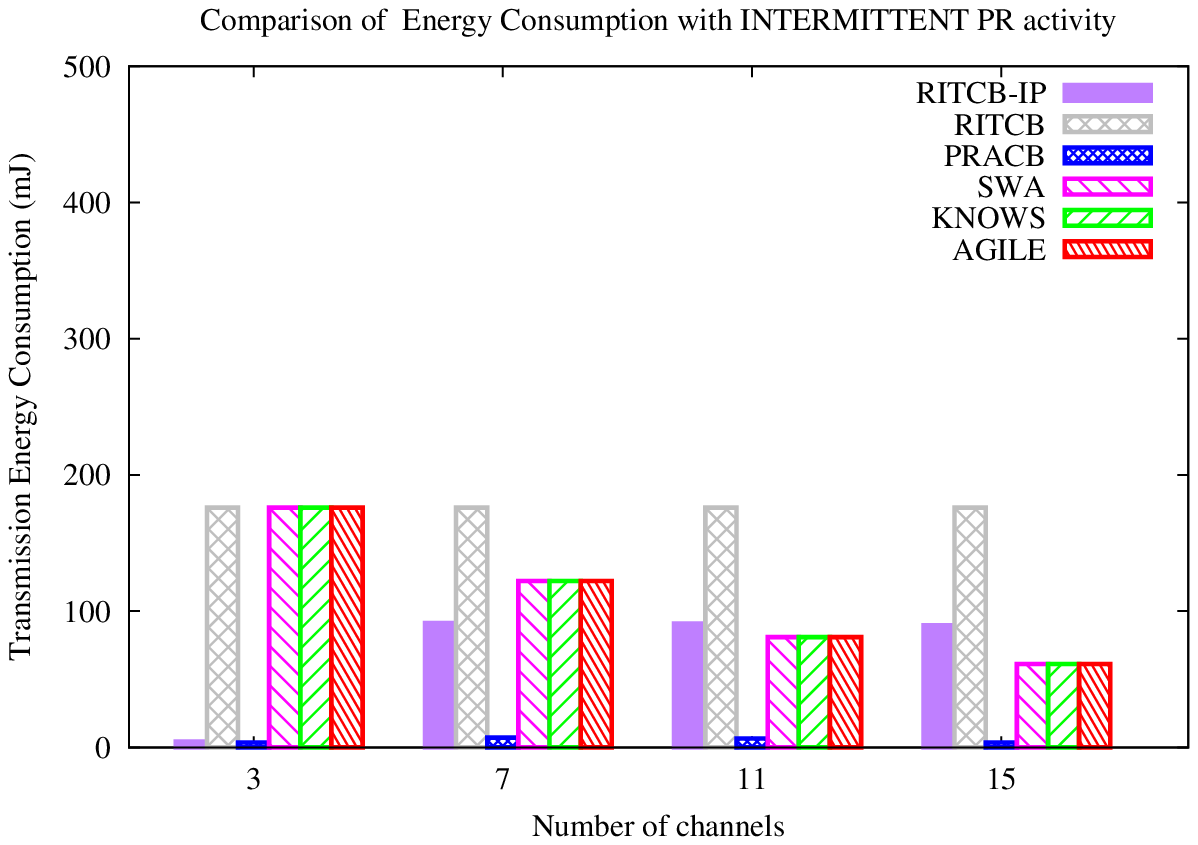}
    }\hspace{0cm}
   \vspace{-0.3cm} \caption{The comparison of CRSN node life time with: (a) Low PR Activity. (b) High PR Activity. (c) Long PR Activity. (d) Intermittent PR Activity.}\vspace{-0.7cm}
    \label{Overview-of-tx-all-PR}
  \end{center}
\end{figure*}

\subsubsection{Node Life Time}

\begin{figure*}[htbp]
  \begin{center}
    \subfigure[]
    {
      \label{RIT-cs-low}
      \epsfxsize= 7cm
	  \leavevmode\epsfbox{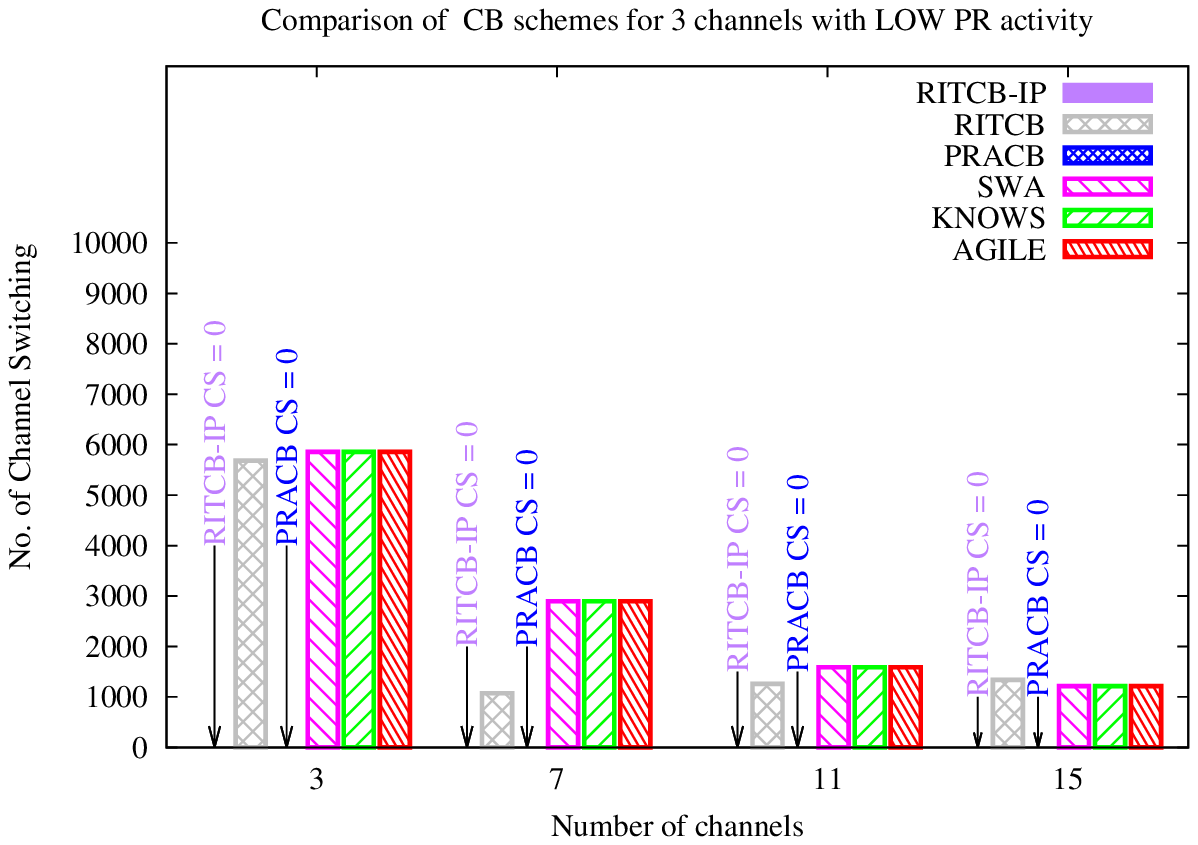}
    }\hspace{0cm}
    \subfigure[]
    {
      \label{RIT-cs-high}
      \epsfxsize= 7cm
	  \leavevmode\epsfbox{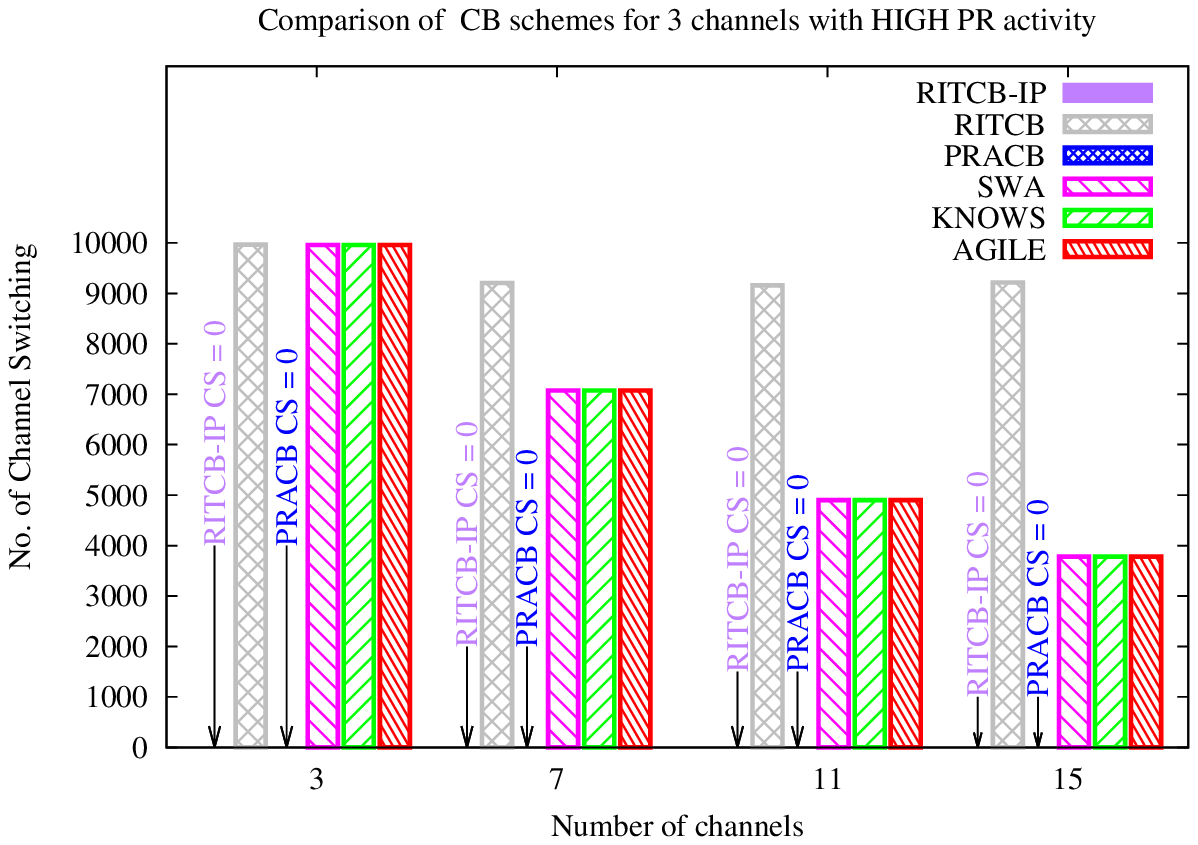}
    }\hspace{0cm}
    \subfigure[]
    {
      \label{RIT-cs-long}
      \epsfxsize= 7cm
	  \leavevmode\epsfbox{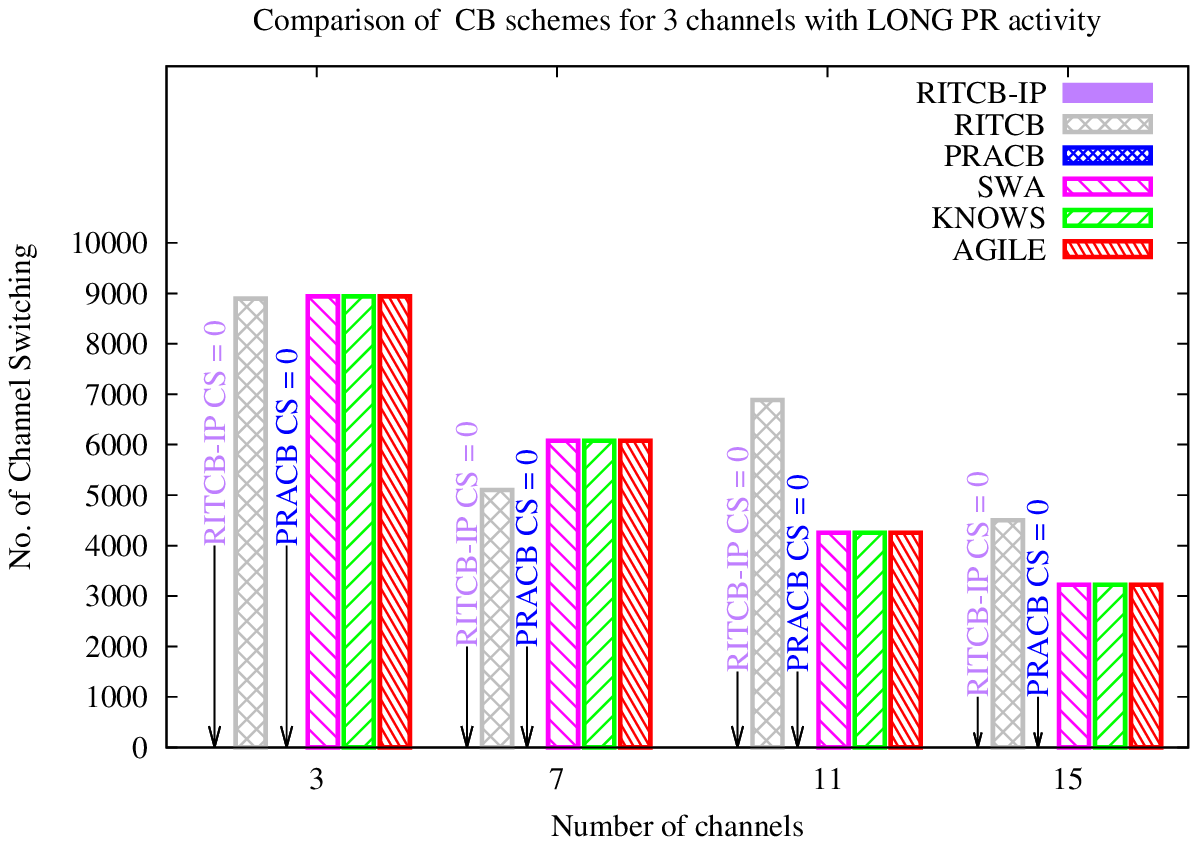}
    }\hspace{0cm}
  \subfigure[]
    {
      \label{RIT-cs-inter}
      \epsfxsize= 7cm
      \leavevmode\epsfbox{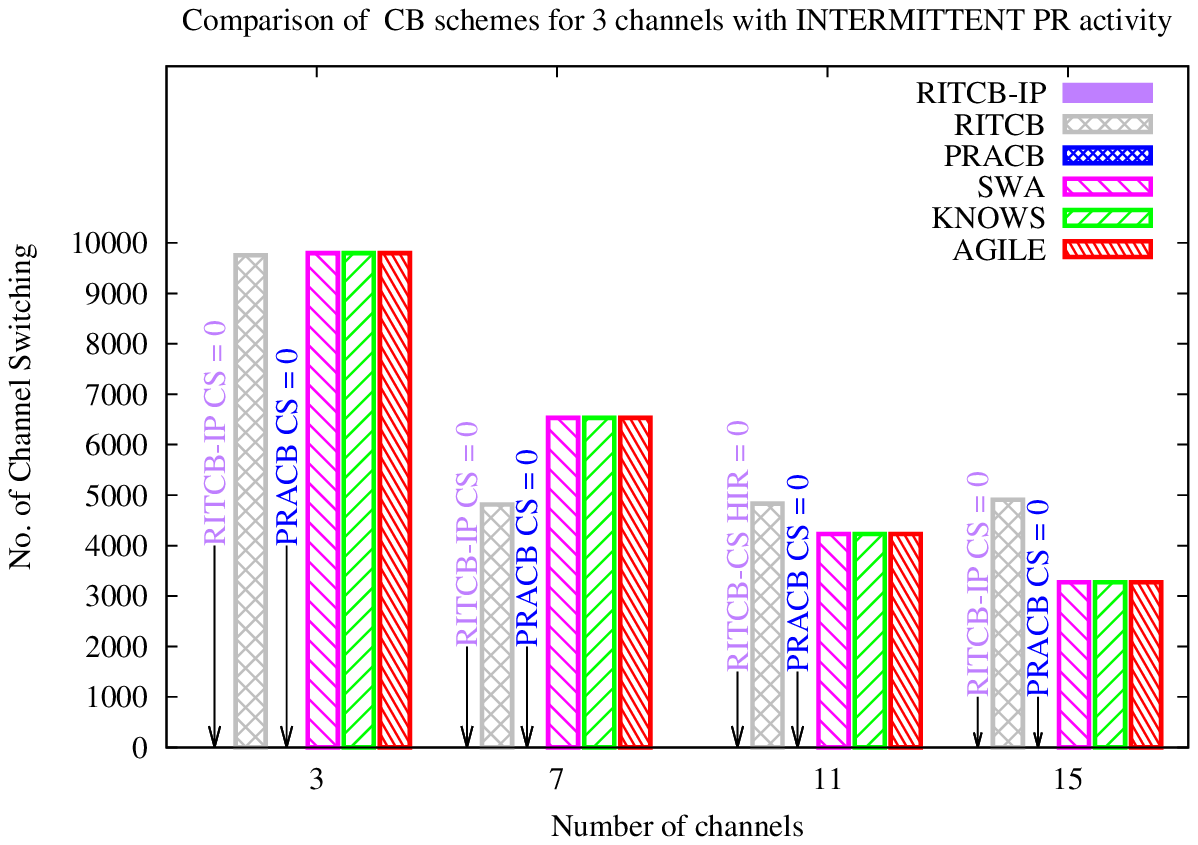}
    }\hspace{0cm}
   \vspace{-0.3cm} \caption{Overview of channel switching between multiple channels with: (a) Low PR Activity. (b) High PR Activity. (c) Long PR Activity. (d) Intermittent PR Activity.}\vspace{-0.7cm}
    \label{Overview-of-cs-all-PR}
  \end{center}
\end{figure*}

The CRSN node life depends upon many factors. The number of total packets transmitted, successfully received packets and the packets which caused harmful interference with PR nodes affect the energy of CRSN node. Lets suppose that initial CRSN node energy is 1J \cite{li2011residual} and transceiver energy consumption is 50nj/bit \cite{Shah2015} then the total energy consumed by CRSN node can be calculated by considering total number of packets successfully transmitted and the number of packets which caused harmful interference. As our proposed schemes RITCB and RITCB-IP select those channels for CB which have longest remaining idle time so we get higher delivery ratio but RITCB causes harmful interference due to which it consumes more energy than RITCB-IP as shown in Fig. \ref{Overview-of-tx-all-PR}.

When there is low PR activity present in the network, it can be observed in Fig. \ref{RIT-tx-low} that RITCB-IP consumes less energy than RITCB but as number of channels is increased, it seems that PRACB performs better RITCB-IP as it gives lower energy consumption. It is due to reason that as number of channels is increased, PRACB select fewer contiguous channels and its delivery ratio decreases (as depicted in Fig. \ref{RIT-dr-low}) which consequently consumes less energy. However, RITCB provides better delivery ratio even with increased number of channels.

Similarly, when high PR activity is present, there is very less chance of obtaining idle contiguous channels due to which RITCB-IP consumes very less energy in packets transmission (as shown in Fig. \ref{RIT-tx-high}) where as RITCB causes harmful interference and consumes more energy. PRACB also consumes very less energy as it does not obtain contiguous pairs as number of channels is increased.


\textit{Hence, it can be concluded that our objective is to optimize the CRSN node life by increasing total number of successful packet transmissions and reducing harmful interference. In this manner, RITCB-IP outperforms all other schemes as it reduces harmful interference to zero and provides better successful packet transmission. \ans{Our proposed scheme RITCB-IP consumes less energy and its specific design for wireless sensor nodes made it suitable for CRSNs.}}

\subsubsection{Channel Switching}

Channel switching is the number of times a CRSN node has to shift to other channels when PR node appears on the current channel. When PR node appears on the selected channel, CRSN node has to stop its transmission and leave the channel immediately for PR node. Then it searches for another available channel and upon finding that channel, it occupies that. This phenomena becomes more complex in case of CB, where CRSN node has to find multiple contiguous channels for communication. When PR node appears on any of that channel, the CRSN node has to break the bond, leave the channels and find other suitable channels.

There are two general types of channel switching: proactive channel switching and reactive channel switching. By using proactive channel switching, a CRSN node has the expectation of appearing a PR node over the selected channel so it can leave the selected channel before PR becomes active. Whereas, in reactive channel switching, the CRSN node leaves the channel upon appearance of PR node over the selected channel.

As shown in Fig. \ref{Overview-of-cs-all-PR}, a CRSN node is sending 10000 packets using CB in presence of PR traffic. RITCB is selecting the channels based on expected value of remaining idle time for each channel but when PR node appear at any instant, it has to perform channel switching. We can also estimate the value of channel switching from harmful interference ratio (Fig. \ref{Overview-of-hir-all-PR}). This is the reason due to which channel switching will very high in high PR activity as shown in Fig. \ref{RIT-cs-high}. The RITCB-IP and PRACB do not transmit their packet when the selected channel has active PR node. The other schemes such as SWA, KNOWS and AGILE do not incorporate the concept of channel switching hence they will be causing harmful interference to PR nodes and reduced delivery ratio. In this paper, proposing a channel switching scheme is not our objective. However, the capability of channel switching can be added to RITCB and RITCB-IP which will further improve their performance. The enhancement of channel switching capability can be considered as future work.

\textit{Hence, channel switching can help to reduce harmful interference and improve delivery ratio but frequent channel switching affects the sensor node energy.}

%
%

\section{Conclusion and Future Work}

In this paper, we have proposed intelligent schemes for channel bonding RITCB and RITCB-IP which select channels based on remaining idle time. RITCB and RITCB-IP have shown significant performance gains in delivery ratio while RITCB-IP outperform RITCB, SWA, KNOWS and AGILE in HIR, transmission energy consumption and required number of channel switching. It improves the overall performance of network as it decreases node energy consumption and increases CRSN node life time. The concept of remaining idle time has its significance in cognitive networks where CR nodes attempt to access the channel opportunistically. The concept has greater impact when CRSN nodes are involved as the life time of battery oriented nodes must be optimized. This paper is a significant contribution to perform efficient channel selection for CRSN nodes. 

\label{conclusion}

\bibliographystyle{IEEEtran}
\bibliography{referencestrans}





\end{document}